%% file: ms_arxiv.tex
\newcommand\Msun{\hbox{M$_\odot$}}
\newcommand\msun{\hbox{M$_\odot$}}
\newcommand\Zsun{\hbox{Z$_\odot$}}
\newcommand\kms{\hbox{$\,$km$\,$s$^{-1}$}}

\newcommand\one{\,{\sc i}}
\newcommand\two{\,{\sc ii}}
\newcommand\three{\,{\sc iii}}

\newcommand\tmult{\multicolumn{2}{c}}
\newcommand\hst{\textit{HST}}
\newcommand\chan{\textit{Chandra}}
\newcommand\spit{\textit{Spitzer}}
\newcommand\twomass{{2MASS}}
\newcommand\ie{\textit{i.\,e.}}
\newcommand\eg{e.\,g.}
\newcommand\cf{\textit{cf.}}
\newcommand\etal{et~al.}

\newcommand\bb{$B_{435}$}
\newcommand\vb{$V_{606}$}
\newcommand\ib{$I_{814}$}
\newcommand\bvi{\textit{BVI}}
\newcommand\bmr{\textit{$B-R$}}

\newcommand\lumin{erg~s$^{-1}$}
\newcommand\HI{H~{\sc i}}
\newcommand\flux{erg~cm$^{-2}$~s$^{-1}$}

\newcommand\airac{$\alpha_\textup{\scriptsize IRAC}$}

\newcommand{\asec}{$^{\prime\prime}$}
\newcommand{\farcs}{\mbox{\ensuremath{.\!\!^{\prime\prime}}}}% fractional arcsecond symbol
\newcommand{\altcite}{\citealt}

\newcommand{\myemail}{iraklis@astro.psu.edu}

\documentclass[preprint2]{aastex}
\usepackage{graphicx,epstopdf,rotating}
 
\shorttitle{Hickson Compact Group 59}
\shortauthors{Konstantopoulos \etal}

\begin{document}

\title{The merger history, AGN and dwarf galaxies of Hickson~Compact~Group~59}%\footnote{Based on observations made with the NASA/ESA Hubble Space Telescope}}

\author{I.~S.~Konstantopoulos\altaffilmark{1},
S.~C.~Gallagher\altaffilmark{2},
K.~Fedotov\altaffilmark{2,3},
P.~R.~Durrell\altaffilmark{4},
P.~Tzanavaris\altaffilmark{5,6,7},
A.~R.~Hill\altaffilmark{2},
A.~I.~Zabludoff\altaffilmark{8},
M.~L.~Maier\altaffilmark{9},
D.~M.~Elmegreen\altaffilmark{10},
J.~C.~Charlton\altaffilmark{1},
K.~E.~Johnson\altaffilmark{11,12},
W.~N.~Brandt\altaffilmark{1},
L.~M.~Walker\altaffilmark{11},
M.~Eracleous\altaffilmark{1},
A.~Maybhate\altaffilmark{13},
C.~Gronwall\altaffilmark{1},
J.~English\altaffilmark{14},
A.~E.~Hornschemeier\altaffilmark{5},
J.~S.~Mulchaey\altaffilmark{15}
}

 \altaffiltext{1}{Department of Astronomy \& Astrophysics, The Pennsylvania State University, University Park, PA 16802; \mbox{\myemail}}
 \altaffiltext{2}{Department of Physics \& Astronomy, The University of Western Ontario, London, ON~N6A~3K7, Canada} 
\altaffiltext{3}{Herzberg Institute of Astrophysics, Victoria, BC~V9E~2E7, Canada}
 \altaffiltext{4}{Department of Physics \& Astronomy, Youngstown State University, Youngstown, OH~44555}
 \altaffiltext{5}{Laboratory for X-ray Astrophysics, NASA Goddard Space Flight Center, Greenbelt, MD~20771}
 \altaffiltext{6}{Department of Physics and Astronomy, The Johns Hopkins University, Baltimore, MD~21218}
 \altaffiltext{7}{NASA Post-doctoral Program Fellow}
 \altaffiltext{8}{Steward Observatory, University of Arizona, Tucson, AZ~85721}
 \altaffiltext{9}{Gemini Observatory, Casilla 603, Colina el Pino S/N, La Serena, Chile}
 \altaffiltext{10}{Department of Physics \& Astronomy, Vassar College, Poughkeepsie, NY~12604}
 \altaffiltext{11}{Department of Astronomy, University of Virginia, Charlottesville, VA~22904}
 \altaffiltext{12}{National Radio Astronomy Observatory, Charlottesville, VA~22903}
 \altaffiltext{13}{Space Telescope Science Institute, Baltimore, MD~21218}
 \altaffiltext{14}{University of Manitoba, Winnipeg, MN, Canada}
 \altaffiltext{15}{Carnegie Observatories, Pasadena, CA ~91101}

\begin{abstract} 

Compact group galaxies often appear unaffected by their unusually dense 
environment. Closer examination can, however, reveal the subtle, cumulative 
effects of multiple galaxy interactions. Hickson Compact Group~(HCG)~59 
is an excellent example of this situation.  We present a photometric study 
of this group in the optical (\hst), infrared (\spit) and X-ray (\chan)
regimes aimed at characterizing the star formation and nuclear
activity in its constituent galaxies and intra-group medium.  We
associate five dwarf galaxies with the group and update the velocity
dispersion, leading to an increase in the dynamical mass of the group of 
up to a factor of 10 (to $2.8\times10^{13}~$\Msun), and a subsequent 
revision of its evolutionary stage.  
Star formation is proceeding at a level consistent with the morphological
types of the four main galaxies, of which two are star-forming and the
other two quiescent. Unlike in some other compact groups, star-forming
complexes across HCG~59 closely follow mass-radius scaling relations
typical of nearby galaxies.  In contrast, the ancient globular cluster
populations in galaxies HCG~59A and B show intriguing irregularities, and two
extragalactic H~{\sc ii} regions are found just west of B.  We age-date
a faint stellar stream in the intra-group medium at $\sim1~$Gyr to examine recent
interactions.  We detect a likely low-luminosity AGN in HCG~59A by its
$\sim10^{40}$~\lumin\ X-ray emission; the active nucleus rather than star
formation can account for the UV+IR SED.  We discuss the implications
of our findings in the context of galaxy evolution in dense
environments.  \end{abstract}

\keywords{galaxies: clusters: individual (HCG~59) --- galaxies: star clusters --- 
galaxies: evolution --- galaxies: interactions --- galaxies: active --- galaxies: dwarf --- galaxies: fundamental parameters}

%======= 1. INTRODUCTION =======
\section{Introduction}\label{sec:intro}

Compact galaxy groups  populate the high density tail of the
galaxy number density distribution. The systems catalogued by
\citet[][Hickson Compact Groups, or HCGs]{hickson82} exhibit some
features, such as dynamical and evolutionary states, elliptical
fractions and X-ray properties of the intra-group medium (IGM) similar to
galaxy clusters, their massive, more populous counterparts.  In contrast to
the well-studied cluster galaxies, however, the specific effects of
the compact group environment on the evolution of its galaxies are not
yet clear.

HCGs are defined through criteria of isolation and surface
brightness\footnote{ $\theta_N\geq 3~\theta_G$, \ie\ a circular area
defined by three galaxy-mean-radii about the group is devoid of galaxies of
comparable brightness.  A group surface brightness of
$\mu<26.0~$mag defines galaxy density.}  that give rise to
self-gravitating, dense groupings of a few (typically four) main
members. Because of their masses, these galaxies orbit around the
group barycenter rather sluggishly, with velocity dispersions on the
order of $\sigma_\textup{\scriptsize CG}\sim250$~\kms\
\citep{tago08,asqu}, \cf\ galaxy cluster dispersions of
$\sigma_\textup{\scriptsize cluster}\sim750~$\kms\
\citep{binggeli87,the86,asqu}.  This trait makes HCGs valuable
laboratories for galaxy evolution: the low velocity dispersions force
some galaxies into strong, prolonged interactions while others appear
undisturbed but are apparently undergoing enhanced secular
evolution. That is to say, this latter population is affected by
gravitational interplay with their neighbors, but evolve more subtly,
without obvious, strong interactions \citep{isk10}.

Relating the various observational characteristics of compact groups
to those of clusters is important for understanding whether they
constitute their own class, or if they are simply
mini-clusters. Perhaps more appropriately, structures like compact
groups may be considered plausible building blocks of clusters at
higher $z$ \citep[\eg][]{fujita04,rudick06}.  
Revealing past investigations of HCGs as a class have
focussed on gas content. Their members are typically deficient in \HI\
gas when compared to galaxies of similar morphological types and
masses \citep[\eg\ the sample of isolated galaxies
in][]{haynes84}. \citet{vm01} proposed an evolutionary sequence based
on mapping the spatial distribution of H\one\ across a large sample of
HCGs.  \citet[][hereafter J07]{johnson07} added to this investigation by quantifying
the gas richness of twelve groups with the relation of
H\one-to-dynamical mass, $\log(M_\textup{\scriptsize
H\one})/\log(M_\textup{\scriptsize dyn}$).  This gave rise to the
hypothesis of an alternate, two-pronged evolutionary diagram for HCGs,
which we explored in \citet{isk10}.  In one path, the galaxies have
strong interactions before exhausting their cold gas reservoirs for
star formation, in the other, gas is processed by star formation
within individual galaxies prior to late-stage dry mergers.

Furthermore, the mid-IR colors of HCG galaxies show an interesting
bimodal distribution that distinguishes star-forming from quiescent
systems. \citet{walker10} interpret this statistically significant gap
as evidence for accelerated galaxy evolution in the compact group
environment. Their similar mid-IR color distributions relate HCGs to
the infall regions of clusters and set them apart from any other
galaxy sample compared, interacting or quiescent.  This theme was
expanded by \citet{tzanavaris10} who found this gap apparent also in
the distribution of specific star formation rates for HCG galaxies.
These observations together point to compact groups as local examples
of the plausible building blocks of clusters in the early universe.

In addition, HCGs, which are isolated by selection, could potentially
help explain the evolutionary history of some field ellipticals. For
example, \citet{rubin90} originally proposed (see also
\altcite{gallagher08}) that HCG~31 will evolve into a single, field
elliptical through a wet merger (one where gas is still available
during the interaction). `Fossil groups', the {probable} ultimate fate of
isolated groupings, were examined by \citet{jones03}, who defined a
criterion of diffuse X-ray emission in excess of
$0.5\times10^{42}h_{70}^{-2}~$erg~s$^{-1}$ for such a
classification. This arises from the processing of a group's IGM
during a merger (or series of mergers), but the low total mass of most
local compact groups suggests their potential well lacks the depth
required to heat the IGM to X-ray detectable levels \citep{mz98}. 
%build up such an X-ray-emitting IGM. 
Using multiple mergers as a
vehicle toward a fossil group end-state maps one path of galaxy
evolution from the `blue cloud' of star-forming disk galaxies to the
`red sequence' of quiescent bulge-dominated galaxies \citep{bell04}.

Fossil group formation may provide an analogy to cluster centers or 
sub-clumps where the buildup of cD galaxies occurs. If this turns 
out to be valid, the study of compact groups could also help illuminate morphological transformations in the innermost cores of clusters. 
Exploring these different scenarios may prove fruitful for our understanding 
of galaxy evolution and the buildup of stellar mass in the universe. 
Making meaningful progress in this area requires detailed multi-wavelength
studies in order to map the range of physical processes affecting
galaxies that are found in these environments, determine their
histories, and project their evolution.

A consistent treatment of a large sample of HCGs is therefore in
order. In this work we continue the series of \citet{gallagher10} and
\citet{isk10} and provide a comprehensive, multi-wavelength study of
HCG~59.  We will look at the current state of the group through its
star formation and nuclear activity; investigate its past through the
star cluster populations; try to unravel the history of mergers in the
group; examine its dwarf galaxy system; and place it in the context of
HCGs in general.

The core of HCG~59 consists of four giant galaxies, {a typical number for
HCGs in general}.  The group lies at a distance of 60~Mpc, based on a
recession velocity of $v_R=4047~$\kms\ \citep[][corrected to the
reference frame defined by the 3K Microwave Background]{hickson92} and
$H_0=73~$\kms\,Mpc$^{-1}$. Three of the galaxies, A~(type Sa), B~(E0), 
and~C~(Sc), have seemingly undisturbed morphologies, and the fourth~(D, Im) 
is an unusually large irregular with a normal, peaked light profile. The 
total stellar mass of the group is \mbox{$M^{*}_\textrm{\scriptsize
TOT}=3.14\times10^{10}~$\Msun} \citep[from the 2MASS $K_s$-band
luminosities;][]{tzanavaris10}, while the H\one\ mass of
\mbox{$M_{\textup{\scriptsize H\one}}=3.09\times10^{9}~$\Msun} is
comparable to the value expected for the morphological types and
stellar masses of the member galaxies, according to \citet{vm01}. This
is therefore a somewhat gas-rich compact group, given that HCGs typically
contain only about a third of the H\one\ expected. On the other hand,
the J07 scheme classifies the H\one\ content of the
galaxy group as a Type II, \ie\ intermediate in gas content, according
to its ratio of gas-to-dynamical mass of $\log(M_\textup{\scriptsize
H\one})/\log(M_\textup{\scriptsize dyn}) = 0.81\pm 0.05$.  These
classifications are based on different criteria and the disparity can
thus be reconciled.  Table~\ref{tab1} summarizes some of the general
characteristics of the four galaxies, while Table~\ref{tab2_dlx} presents
some derived and literature values of the mass content and nuclear
identifications in the four galaxies.

This paper is organized in the following way: Section~\ref{sec:obs}
presents the optical, IR, and X-ray datasets used throughout this
work. Section~\ref{sec:clusters} provides a full account of
 the young and old star cluster populations, which we use as our prime
 diagnostics of current
 star formation and ancient interactions. In Section~\ref{sec:results}
 we discuss the main findings
 of this work. Finally, in Section~\ref{sec:summary} we summarize the
 work presented and offer ties
 to previous and future work in this series.

%%%--- TABLE 1: BASIC INFO ---%%%
\input{tab1}
%%%

%%%--- Figure 1: Pretty pictures ---%%%
\begin{figure*}[htbp]
\begin{center}

	\includegraphics[width=\textwidth, angle=0]{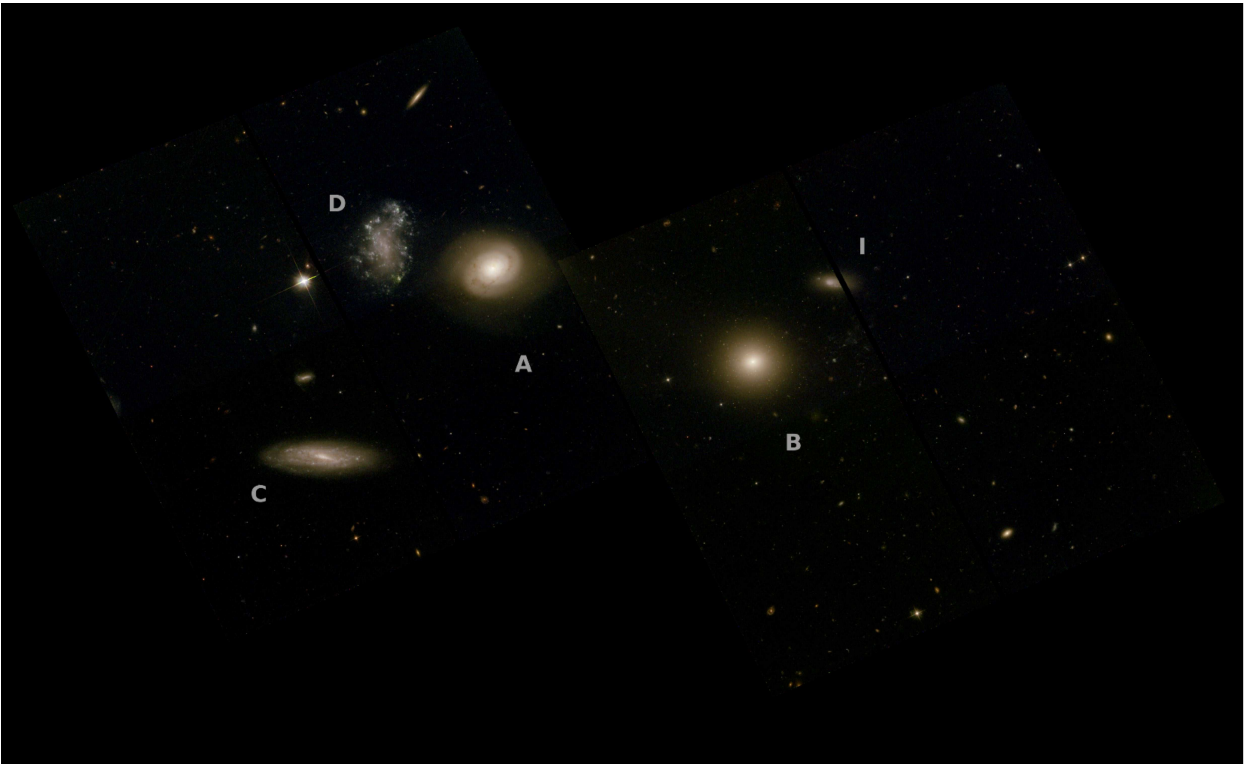}
	\caption{
		\hst~\textit{BVI} color-composite imaging of HCG~59. 
		The red colors of galaxies~A~and~B imply little star formation, while 
		C~and~D show some signatures of nebular emission in green. Also 
		visible is the newly catalogued dwarf galaxy~I (Section~\ref{sec:dwarfs}). 
	}\label{fig:finder-hst}
	
\end{center}
\end{figure*}

%%%--- TABLE 2: MASS, SFR, NUCLEI ---%%%
\input{tab2}
%%%

%======= OBSERVATIONS =======
\section{Observations}\label{sec:obs}
\subsection{\hst\ optical imaging}\label{sec:obs-hst} 

The analysis presented in this paper is based largely on \hst-ACS/WFC
multi-band data. Images were taken in the \textit{F435W, F606W
\textup{and} F814W} bands in two pointings to cover all known giant
group members. We will refer to these filters as \bb, \vb, \ib\ (and
the set as \textit{BVI}) to denote the closest matches in the Johnson
photometric system. The notation does not, however, imply a conversion
between the two systems.  The observations were executed on 2007~February~24,
as part of GO program 10787 (PI: Jane Charlton). The exposure times
were 1710,~1230~and~1065~seconds in the \textit{BVI} bands
respectively.  Three equal sub-exposures were taken with each filter
with a three-point dither pattern (sub-pixel dithering). Images were
reduced `on the fly' to produce combined, geometrically corrected,
cosmic-ray cleaned images. For the analysis of point sources, we used
the standard \hst\ pipeline products with a nominal pixel scale of
$0\farcs05$ per pixel. For analysis of the extended sources, we ran
\texttt{MultiDrizzle} \citep{multidrizzle} with the pixel scale set to
$0\farcs03$ per pixel to improve the spatial resolution.  The absolute
image astrometry was checked with the world coordinate system of the
Two Micron All-Sky Survey catalog \citep[2MASS;][]{2mass} by
identifying four unsaturated point sources in common; the average
offset was $\sim0.01$\arcsec\ in RA and Dec.  The four main galaxy
\ib-band light profiles were fit with S\'{e}rsic profiles using GALFIT
\citep{sersic,peng}; the best-fitting centroid positions are given in
Table~\ref{tab1}.

We used the images, presented in Figure~\ref{fig:finder-hst}, to
characterize the optical morphology of the galaxies, and to detect and
photometer star clusters and cluster complexes.  All reported
magnitudes are in the Vega magnitude system.  In
Section~\ref{sec:clusters}, we present the analysis of these two
scales (clusters and complexes) of the star formation hierarchy and also distinguish between
young massive clusters~(YMCs) and globular clusters~(GCs).

\subsection{Optical point source photometry}\label{sec:sc-phot} We
follow the same rationale applied in our previous work
\citep{gallagher10,isk10} and use star clusters to infer the star
formation activity and history in each of the HCG~59 galaxies.  At the
adopted distance to HCG~59 of 60~Mpc, we expect some contamination by
supergiant stars, which can have absolute $V$-band magnitudes as
bright as $-8.5$~\citep{efremov86}.  At this distance, one ACS pixel
measures $\sim13$~pc, \citep[\cf\ the average star cluster radius of
$\sim4$~pc; \eg][]{remco07}, meaning that clusters are at most
marginally resolved and can be considered point sources for the
purposes of selection and photometry.  We select clusters using the
method described in \citet{gallagher10}; in brief, we perform the
initial selection on median-divided images, require selection in all
three bands, and filter the resulting catalog using point spread
function (PSF) photometry.  Our PSF filtering applied the following
criteria from the output of the \texttt{ALLSTAR} routine in
IRAF\footnote{
	IRAF is distributed by the National Optical Astronomy Observatories,
	which are operated by the Association of Universities for Research
	in Astronomy, Inc., under cooperative agreement with the National
	Science Foundation.}: 
$\chi$ values below 3.0; a sharpness in the range $[-2.0,2.0]$; and a
photometric error less than $0.3~$mag. Aperture corrections are first
measured between 3 and 10 pixels and then added to the
\citet{sirianni05} corrections to infinity.  Finally, foreground
(Galactic) extinction with $E(B-V)=0.037$ is accounted for using the
standard Galactic extinction law \citep[a correction of
$A_V\sim0.12$~mag; ][]{schlegel98}.

In order to fortify the selection against stars, we apply a
conservative absolute magnitude cut at $M_V<-9$~mag, which produces
the high-confidence sample. We do, however, define a larger sample by
relaxing the magnitude cut and applying stricter PSF-fitting
criteria to detect globular clusters, which are expected to be fainter
and point-like. In order to minimize contamination from marginally
resolved sources such as compact background galaxies, we follow
\citet{rejkuba05} and apply hyperbolic filters (starting narrow for
bright sources and widening for fainter sources) with a maximum
cut-off at the above mentioned criteria of magnitude error, $\chi$ and
sharpness.  We will refer to this as the extended sample.

The application of these criteria assigns 240 bright star cluster
candidates (SCCs) to the high-confidence sample and 948 to the
extended sample.  Specifically, the numbers of detected SSCs (extended sample
numbers in parentheses) in galaxies A through D are 7~(29), 77~(213),
13~(63) and 65~(217), with a further 78~(426) objects coincident with
what would be the intra-group medium. We will provide a full analysis of these
cluster populations in Section~\ref{sec:scs}.

In order to test the completeness of the final list of SCCs, we used 
\texttt{ADDSTAR} to add 3000~artificial stars to the image (over the 
entire field, including the galaxies) in the apparent 
magnitude range 24--28, \ie\ absolute magnitudes of $(-9.89, -5.89)$. 
{Because the final catalogue only contains sources detected in all three filters, we include this effect by calculating completeness fractions based only on artificial stars detected in all three bands \citep[\eg][]{darocha02}.}
The limiting magnitudes for the 90\% and 50\% recovery rates are (26.56, 27.25), (26.51, 27.19) and (26.47, 27.15) in the $B_{435}$ , $V_{606}$ and $I_{814}$ bands respectively (after photometric corrections are applied).
% 
%The limiting magnitudes for the 90\%\ and 50\%\ recovery rates are 
%(26.85, 27.45), (26.72, 27.50) and (26.50, 27.12) in the \bb, \vb\ 
%and \ib\ bands respectively (after photometric corrections are applied). 
% 
For the distance modulus used of 33.89~mag, $m=26.5$~mag corresponds to 
$M\simeq-7.4$~mag. 

{Our assessment of the state of star formation in HCG~59 is not limited
to star clusters. Star cluster complexes represent a larger scale of star formation, 
as the optically blended concentrations of gas, stars, and dust that make up small 
star-forming regions, and likely include groups of clusters. In contrast to star clusters, these can be resolved to even 
greater distances than studied here, as the fractal distribution of gas about a 
galaxy gives rise to such structures at all scales 
\citep[\eg\ as demonstrated for M33 by][]{bastian07}. 
}

\subsection{Globular Cluster Candidate Selection}
\label{sec:gc-selec}

Globular clusters are also selected from the \hst\ images. Since the
process is tuned to the color distributions found in HCG~59, we
provide a full account below.  As contamination from supergiants is
not a problem for objects with GC-like colors, we adopt a fainter
magnitude limit to select old GC candidates than we used for SCCs.  We
have chosen a cutoff at $V_{606}=26$, which corresponds to $M_V\sim
-7.7$ at our adopted distance modulus for HCG 59, or slightly more
luminous than the expected peak in the globular cluster luminosity
function (GCLF) at $M_V\sim -7.4$ \citep[e.g.][]{az98,harris01}.  The
majority of GC candidates brighter than this limit lie above the 90\%
photometric completeness level in all 3 filters.  Assuming a Gaussian
GC luminosity function with a peak at $M_V=-7.4\pm 0.2$ and a
dispersion $\sigma=1.2\pm 0.2$, our faint-end cutoff then samples
$39\pm 8 \%$ of the entire GCLF.

\begin{figure*}[htbp]
\begin{center}
	
	\includegraphics[width=\textwidth]{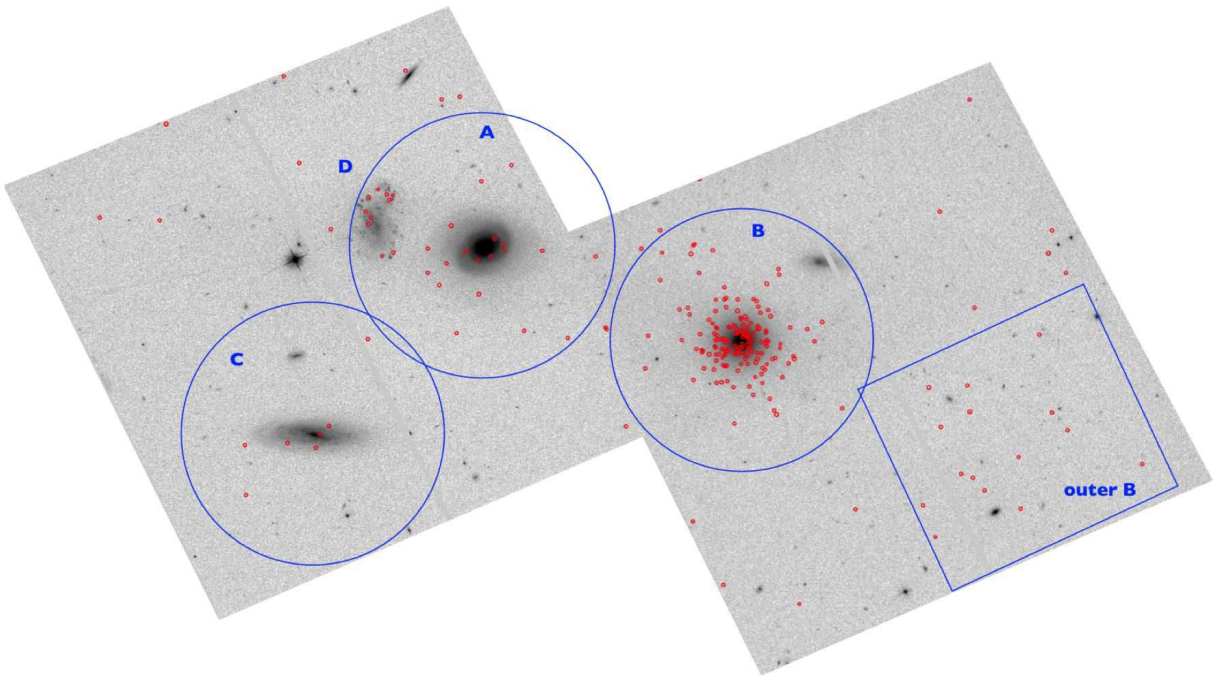}
	
	\caption{Positions of detected globular cluster candidates,
	marked on \hst\
	\ib\ imaging. We have also marked the expected GC haloes with
	blue boundaries.
	`Outer~B' refers to the location of a possible excess of GCs,
	discussed in
	the text. This is symmetric to the `A-B bridge' (with respect
	to B; see Fig.~\ref{fig:bplusr}) and could
	be a tidally induced redistribution of the GC population (see 
	Sections~\ref{sec:gcs}~and~\ref{sec:hcg59b}). 
	}\label{fig:gc-loc}
	
\end{center}
\end{figure*}

We have selected GC candidates (GCCs) according to the color-space
distribution of Milky Way GCs.  We de-reddened the colors of globular
clusters from the \citet{harris96} catalog by their listed $E(B-V)$
values, and then defined a parallelogram based on the intrinsic
$(B-V)-(V-I)$ color distribution of the MW GCs. This parallelogram was then
converted to the ACS filters using the `synthetic' transformations in
\citet{sirianni05}.  The color selection region adopted here is 0.10
mag wider in $(V-I)$ than that used in the analysis of HCG
7~\citep{isk10}, but still does not exceed the boundaries of the
\citet{harris96} MW GCs.  To quantify, 95 of 97 MW GCs from the
\citet{harris96} catalog (those with \bvi\ information) lie within
this box.  All point sources in the HCG 59 fields with 1\,$\sigma$
error bars that overlap our color selection region are considered GC
candidates, and are plotted in Figure~\ref{fig:gc-loc}.

%%%--- Figure 2: GC locations ---%%%

Due to the close (projected) proximity of the galaxies in the group,
it is likely that the halo GCs in each system will appear superposed.
In an attempt to quantify the GCCs in each galaxy, we use the
relationship between the galactic mass and the radial extent of the GC
systems in galaxies of \citet{rhode07}.  To compute the expected size
of each halo, we have adopted the mass-to-light conversions in that
work, although we stress the general conclusions we reach are not
dependent on the detailed size of any given halo.  As the predicted
masses of all of the group galaxies are just below the lowest mass
galaxies in the \citet{rhode07} sample, we adopted a radial extent of
15 kpc (or 56\arcsec\ at the assumed distance to HCG 59) for each of
galaxies~A,~B~and~C. Galaxy~D is a lower luminosity system, and as
seen in Figure~\ref{fig:gc-loc}, GCCs in this object already lie
within the projected halo of GCCs in galaxy~A. Discussion of the
individual GC systems in each group galaxy follows in
Section~\ref{sec:gcs}.

\subsection{Background and Foreground Contamination}

Contamination in our color-selected sample of GC candidates is
expected from a variety of sources, including foreground Milky Way
halo stars, reddened young clusters, and unresolved background
galaxies. With the small number of GC candidates present in some of
the HCG 59 galaxies (discussed below), contamination can be
significant.

Predictions from Milky Way star count models \citep[the Besan\c con
model of][]{robin03} suggest that only 3--4 foreground Milky Way stars will
appear in the magnitude and color ranges for expected GCs in each of our
ACS fields.  Determining the contamination from younger, reddened clusters is
more difficult, particularly in the central regions of the late-type
galaxies~C~and~D, where such objects could be present.  Unresolved
background galaxies are not likely contributing in any significant way
to the numbers of objects in our fields; analyses of the background
objects (with GC-like colors) in HCG~7 \citep{isk10} showed that the
predicted foreground Milky Way star counts were similar to the
observed putative background contamination, leaving little room for
background galaxies to contribute significantly.

{We also consider the \citet{pirzkal05} analysis of stars in the
  Hubble Ultra-Deep Field (HUDF).  Within the range of colors shown in
  Figure~\ref{fig:colors-all}, they found the main contaminant of
  `void sky' to be M-stars, however, with a \vb-\ib\ of $\sim2.0$,
  they are too red to be considered in our analysis. All Main Sequence
  stars detected in the HUDF are too bright to be mistaken for star
  clusters by our detection algorithm. In fact, the only class of
  stellar object that can be found in the color-space occupied by our
  cluster candidates is white dwarfs, which \citeauthor{pirzkal05}
  find to have a density of $1.1\pm0.3 \times 10^{-2}~$pc$^{-3}$. The
  maximum Galactic volume covered by our two pointings is a cube of
  $\sim8000~$px on a side, or $\sim0.24~$pc$^{-3}$, assuming a scale
  height of 400~pc (the maximum height quoted by
  \citeauthor{pirzkal05}). Such a volume might be expected to host
  $\sim2.6\times10^{-3}$ white dwarfs.  Therefore, we consider this
  potential source of contamination to be negligible.}

\begin{figure*}[htbp]
\begin{center}
		
	\includegraphics[width=0.42\textwidth]{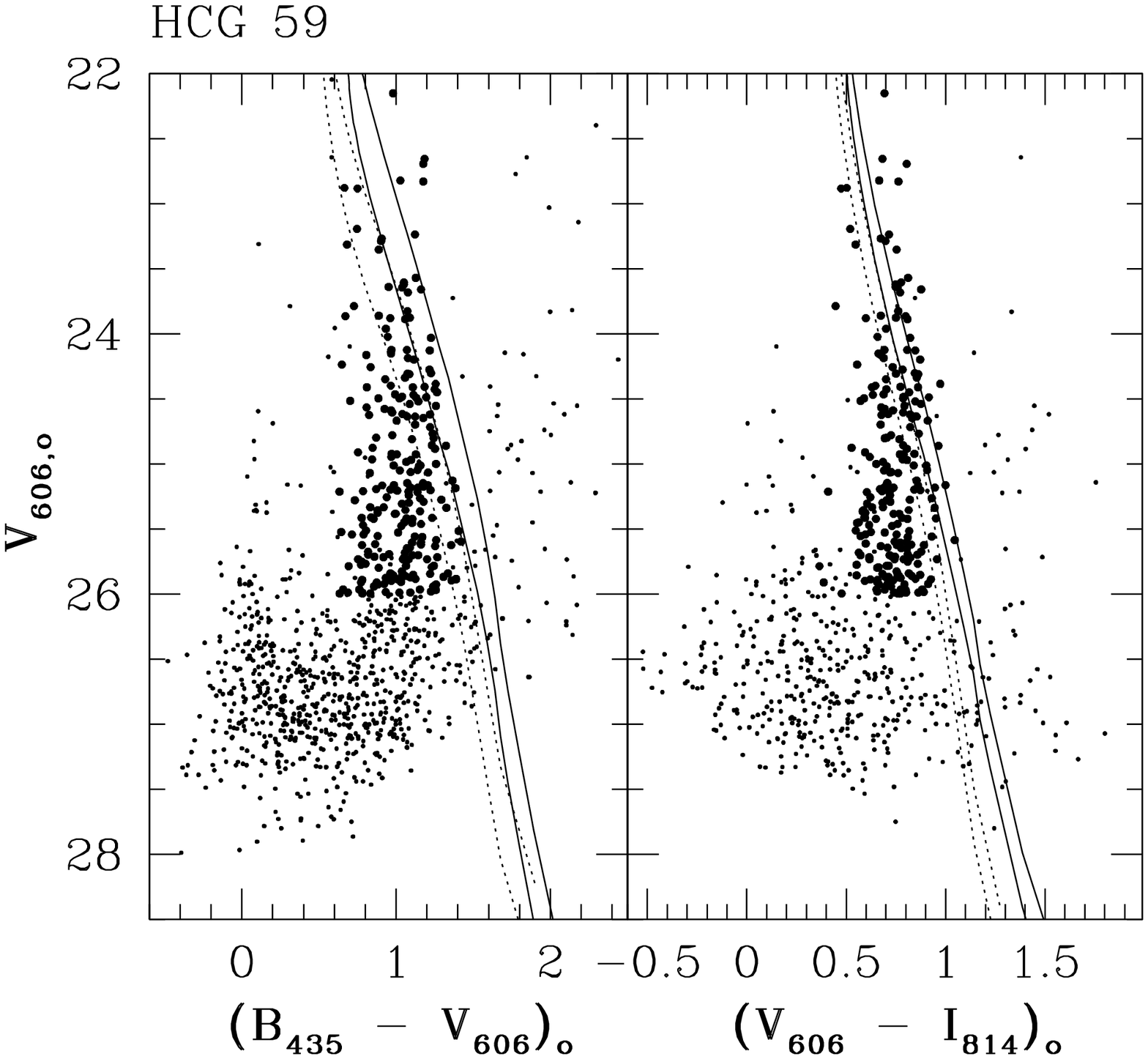}~
	\includegraphics[width=0.45\textwidth]{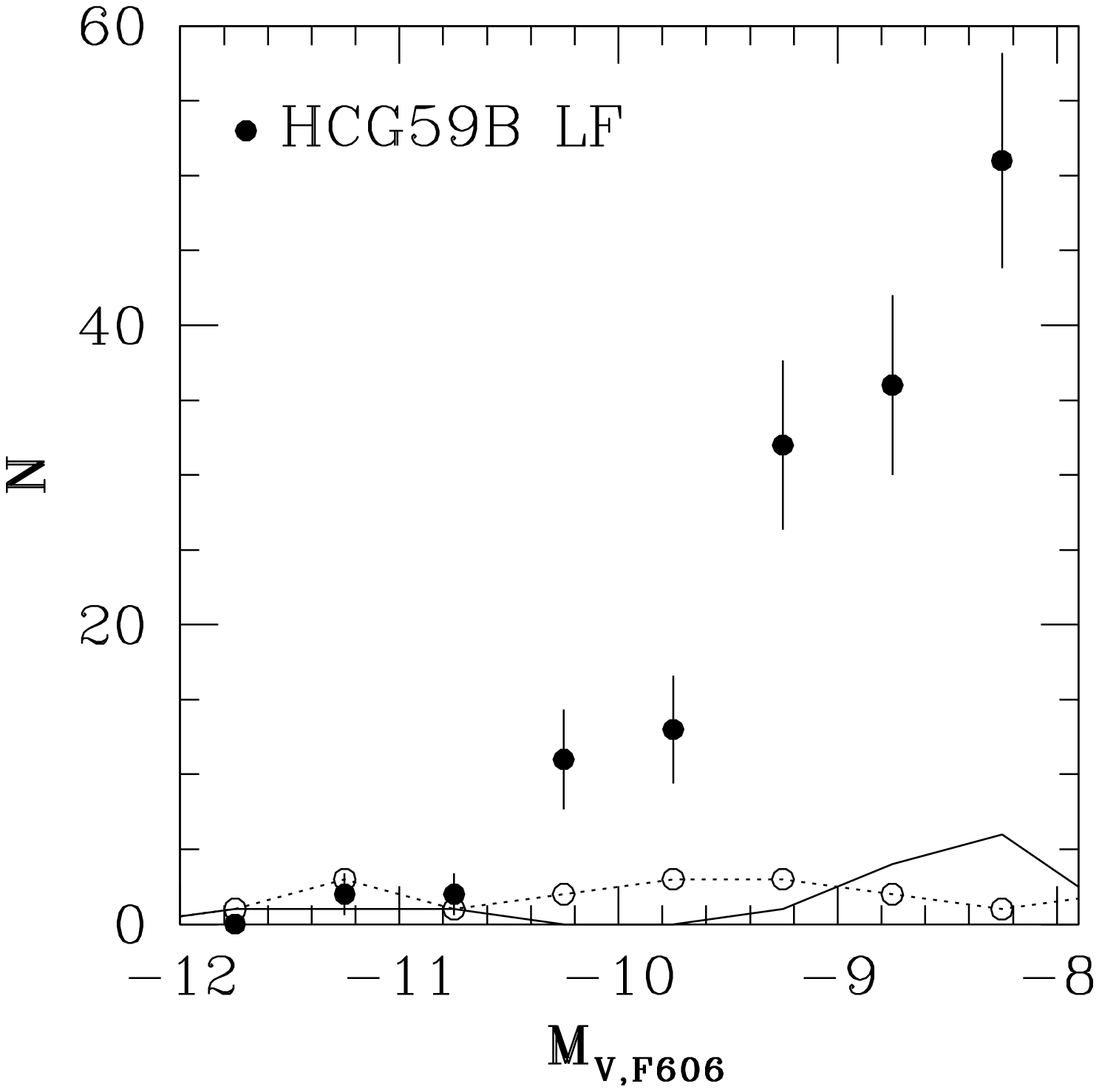}
	\caption{Diagnostic diagrams related to the GC population. 
		On the {\bf left} we show color-magnitude diagrams of
		all detected GCCs.
		The solid and dashed lines show \citet{marigo08}
		evolutionary tracks for
		12~Gyr-old main sequence stars, with $Z=-0.6$ and $
		-1.3$
		respectively. The
		lines are double to bracket the distances appropriate
		for the Sagittarius
		dwarf, with its leading spiral arm superposed along
		this line of sight.
		Their distribution in the CMD dispels our worries of
		significant contamination.
		The {\bf right} panel shows the luminosity function of
		globular clusters candidates (filled circles)
		in galaxy~B and sources considered `background' (open circles and dashed lines).  The clear discrepancy supports the quality of our GC selection. The LF of `outer B' sources is designated with a solid line.  
	}\label{fig:gc-diag}
	
\end{center}
\end{figure*}

The dominant source of contamination to our GCC sample will actually
be stars from the Sagittarius dwarf galaxy.  Our ACS fields (at $l
\sim 254^\circ$, $b\sim +69^\circ$) are superposed on a rather dense part
of the tidally induced (and bifurcated) leading arm \citep[stream `A'
from][]{bel06} of Sgr \citep[e.g.,][]{maj03,bel06,new07,yan09}; the
presence of such streams is not accounted for in traditional Milky Way
star count models.  To investigate the impact of Sgr Stream stellar
populations, in Figure~\ref{fig:gc-diag} we have overlaid 12~Gyr
isochrones of \citet{marigo08} with a range of metallicities expected
for the Sgr leading arm, $[\textup{M/H}]\sim -1 \pm
0.5$ \citep[e.g. ][]{chou2007,yan09}, at distances between 26 and 36
kpc onto color-magnitude diagrams of the point sources in both of our ACS fields
\citep[assuming $d\sim 31$ kpc, with a spread of $\pm 5$
kpc;][]{new07,no10,correnti2010}.  From this, we see that stars just
below the Sgr Stream main sequence turnoff do have colors and
magnitudes similar to that of the brighter ($V_{606} < 25$) GC
candidates in our study, making some contamination likely.

%%--- Figure 3: GC CCs ---%%%

To estimate the {\it total} contamination in our GCC sample, we assume
those GC candidates that lie far outside the GC system halos (as shown
in Figure~\ref{fig:gc-loc}) are instead contaminating sources.  The
one exception to this is a region (called `outer B' in
Figure~\ref{fig:gc-loc}) that lies outside the GC system of galaxy~B,
opposite to the direction of galaxy~A.  We will return to this feature
below.  There are a total of 18 objects in 8.4 arcmin$^2$, or a
background surface density $\Sigma_{back}=2.3 \pm
0.5~\textup{arcmin}^{-2}$.  This is much higher than the predicted
surface density of MW halo stars from the Besan\c con model
($\Sigma_\textup{\scriptsize MW}\sim 0.4~\textup{arcmin}^{-2}$),
indicating that Sgr leading arm stars are the dominant foreground
source of contamination in our sample.

Of course, for the above analysis we are making the assumption that
these contaminating objects are not \textit{bona-fide} `intra-group' GCs that
lie far outside the main galaxies of the group.  To test this, we
compare the $V_{606}$ luminosity function for the background source
sample with the luminosity function of the large GC candidate
population surrounding galaxy~B. We show this in the right panel of
Figure~\ref{fig:gc-diag}.  The extrahalo luminosity function does not
show a sharp rise with increasing magnitude as expected of a GC
luminosity function and exhibited by the GCCs in galaxy~B.  Although a
definitive comparison is not possible with so few objects in the IGM
area, this is consistent with these IGM objects being contaminants and
not a part of a diffuse population of intragroup GCs.  Thus we adopt
the surface density above as that of the `background' in the analyses
that follow.

%======= 

\subsection{Las Campanas wide-field imaging: low surface
brightness light and dwarf galaxies}\label{sec:campanas}

We extend the coverage of the \hst\ observations through wide-field
imaging with the Las~Campanas~Observatory~(LCO)~2.5-meter
telescope. We took $B$- and $R$-band images of a 25\arcmin\ diameter
around the group with the Wide Field Reimaging CCD Camera (WFCCD). The
data were obtained on \mbox{2007~July~07} as part of an imaging
campaign that covers all 12 HCGs in the J07 sample. The
$B$ and $R$ filter exposure times were 300~s and 600~s, respectively.

\begin{figure*}[htbp]
\begin{center}

	\includegraphics[height=\textwidth, angle=270]{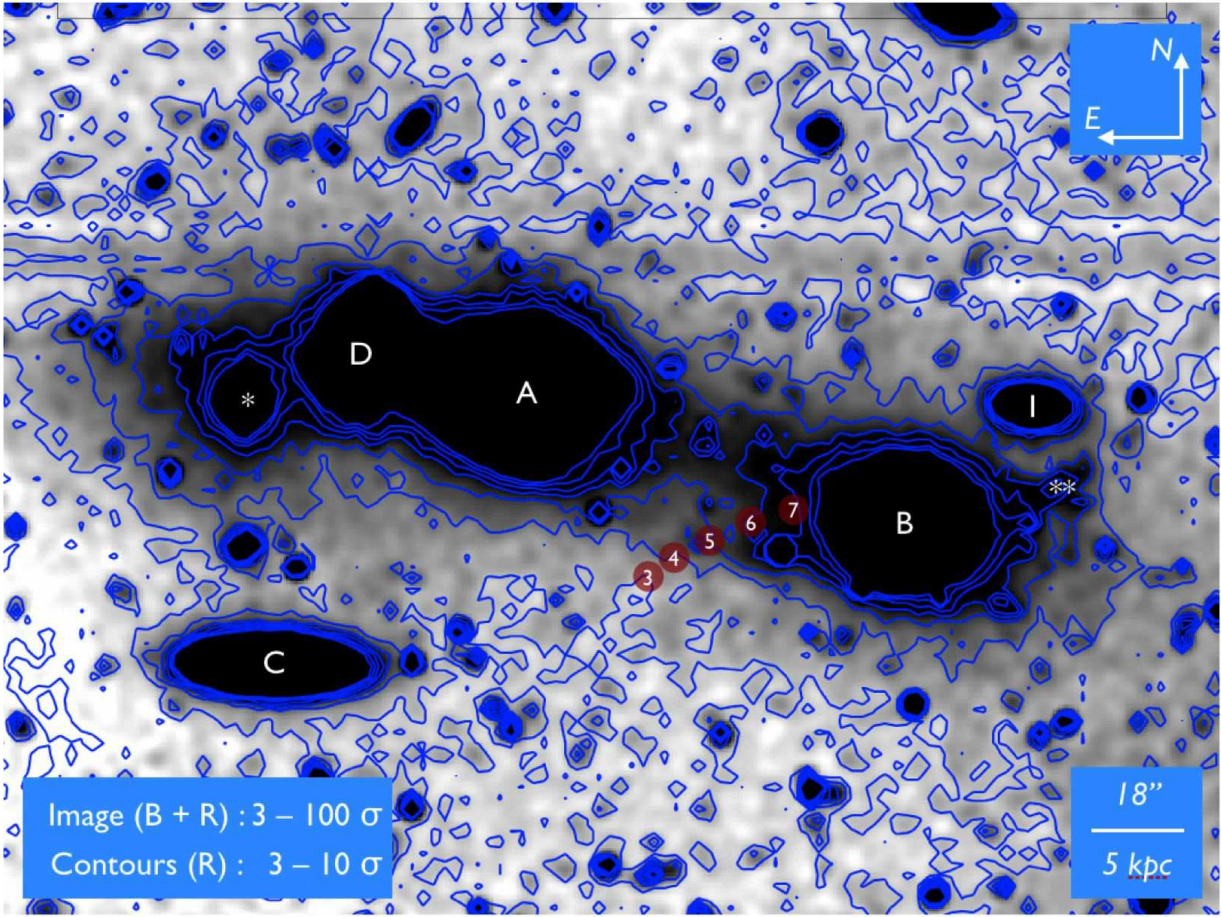}\\
	\caption{Co-added, smoothed, background-suppressed $B+R$ image
	of HCG~59
		from the LCO duPont telescope. The image is aligned to
		the world coordinate
		system. The contours mark \mbox{$R$-band} fluxes
		between
		3~and~10$\,\sigma$ above the background (as marked by
		the circled numbers),
		while the smoothed $B+R$ flux shows material brighter
		than $3\,\sigma$ above the
		background level. The asterisk label denotes the
		foreground star to the east
		of galaxy~D, while the double asterisk marks the
		H\two\ regions west of B.
		The horizontal banding north of the main group is an
		artifact in the
		\mbox{$R$-band} image. 
		We detect a streak of material apparently connecting
		galaxies~A~and~B in our
		field of view. We argue in
		Section~\ref{sec:bplusr} that this is a
		\textit{bona-fide} tidal bridge, although its low
		surface brightness prohibits
		an in-depth photometric study. A faint
		arc of light to the west of galaxy~B -- containing the
		H~{\sc ii} regions mentioned above (see
		\S~\ref{sec:scs}) -- appears to connect it with
		the dwarf galaxy~I. 
		We discuss the importance of these features in
		Section~\ref{sec:bplusr}.
	}
	\label{fig:bplusr}
	
\end{center}
\end{figure*}

These images allow for the detection of low surface brightness
features, such as the signatures of past interactions, over a large
area. We present this analysis in Figure~\ref{fig:bplusr}, where we
stacked the $B$ and $R$ images and applied a Gaussian smoothing filter
to the result. This image shows only features and $R$-band contours
that register at least $3\,\sigma$ above the background.  We find two
faint features, a `bridge' that appears to connect galaxies A~and~B
and an arc extending from B toward a compact structure to its
north-west (we will later refer to this as the `B-I arc'). There is 
another compact, extended source in the space between galaxies C~and~D.

%%%--- Figure 4: Bridge (A, B) ---%%%

The original purpose of the LCO observing program was to prepare a
sample of dwarf galaxy candidates for spectroscopic follow-up. Though
our redshift survey has yet to cover HCG 59, it is covered by the
Sloan Digital Sky Survey~\citep[SDSS;][]{sdss}: a spectroscopic search
sweeping a radius of 30~arcminutes around the nominal center of the
group (the geometric center of the region enveloping the four known
members) yields seven spectra with redshifts in the range 0.01--0.02:
galaxies~C~and~D and five compact galaxies. We therefore consider the
membership of SDSS galaxies~J114817.89+124333.1 and
J114813.50+123919.2, which are covered by our wide-field imaging and
J114930.72+124037.5, J114940.11+122338.6 and J114912.21+123753.8 which
lie at projected distances greater than 13~arcminutes from the group
center. The first of these galaxies is also present in the \hst\
imaging, but lies partly in the ACS chip gap.  In
Section~\ref{sec:dwarfs} we attempt to determine whether these are
HCG~59 members through a phase-space analysis.

The \citet{hickson82} naming convention assigns letters in order of
brightness. Since our imaging does not cover all five dwarf
candidates, we used the SDSS $r$-band photometry to consistently
classify the galaxies as HCG~59~F~through~J. We have omitted the
letter~E, as it was assigned in the original catalog to a background
galaxy. We attempted to measure stellar masses for these galaxies
using 2MASS $K$-band images \citep{2mass}, however, they are below the
detection limit of that survey. Table~\ref{tab:dwarfs} summarizes all
of the information presented in this section: measured and SDSS
photometry, radial velocities, galaxy morphologies and projected
distances from the group barycenter. The latter two properties will be
discussed in Section~\ref{sec:dwarfs}

%%%--- TABLE: DWARF GALAXIES ---%%%
\input{tab3}

% Spitzer -----------
\subsection{\spit\ observations: infrared spectral energy distributions}\label{sec:obs-spit}

\begin{figure*}[htbp]
%\begin{center}

	\includegraphics[width=0.6\textwidth]{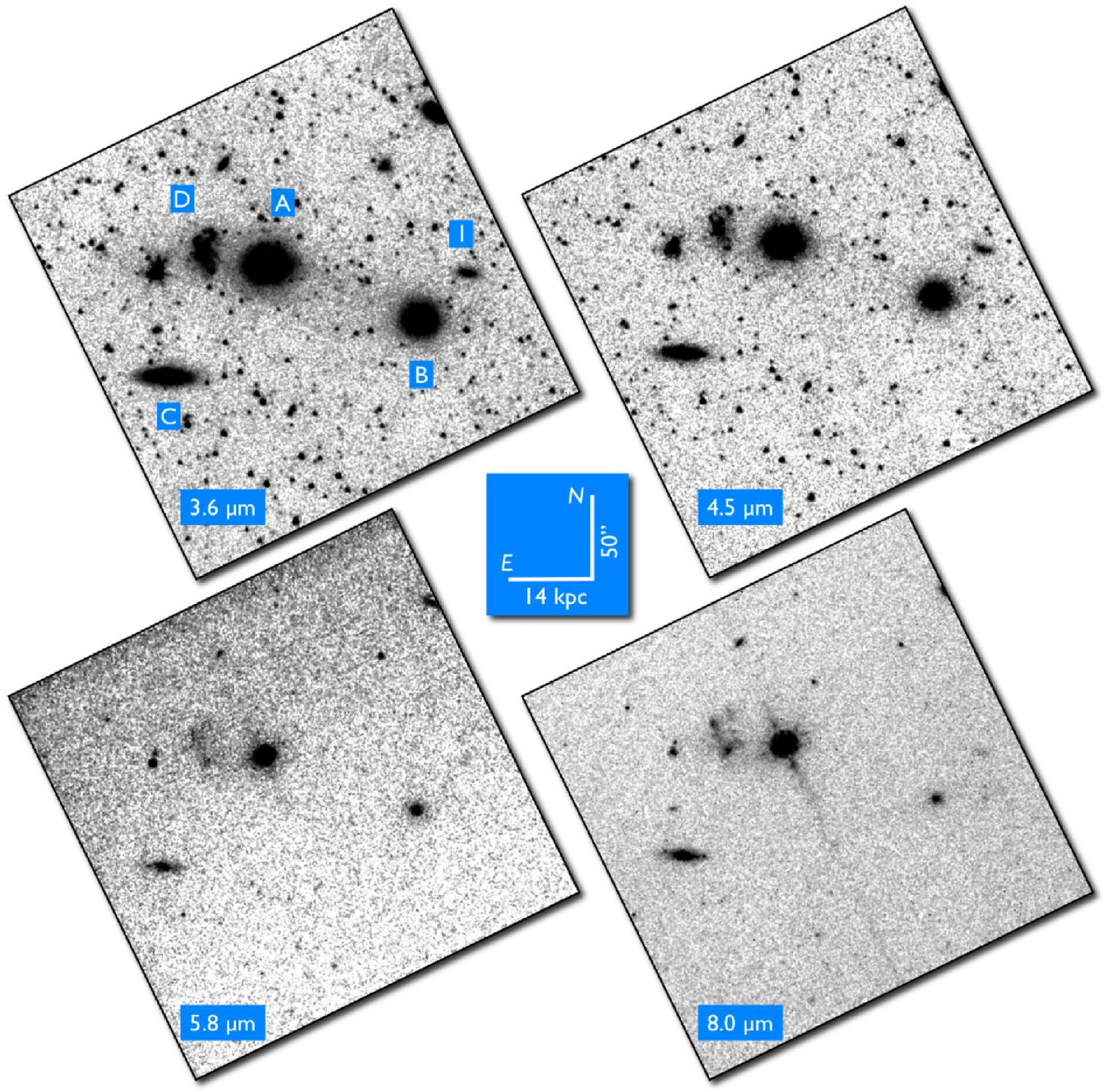}
	\includegraphics[width=0.5\textwidth]{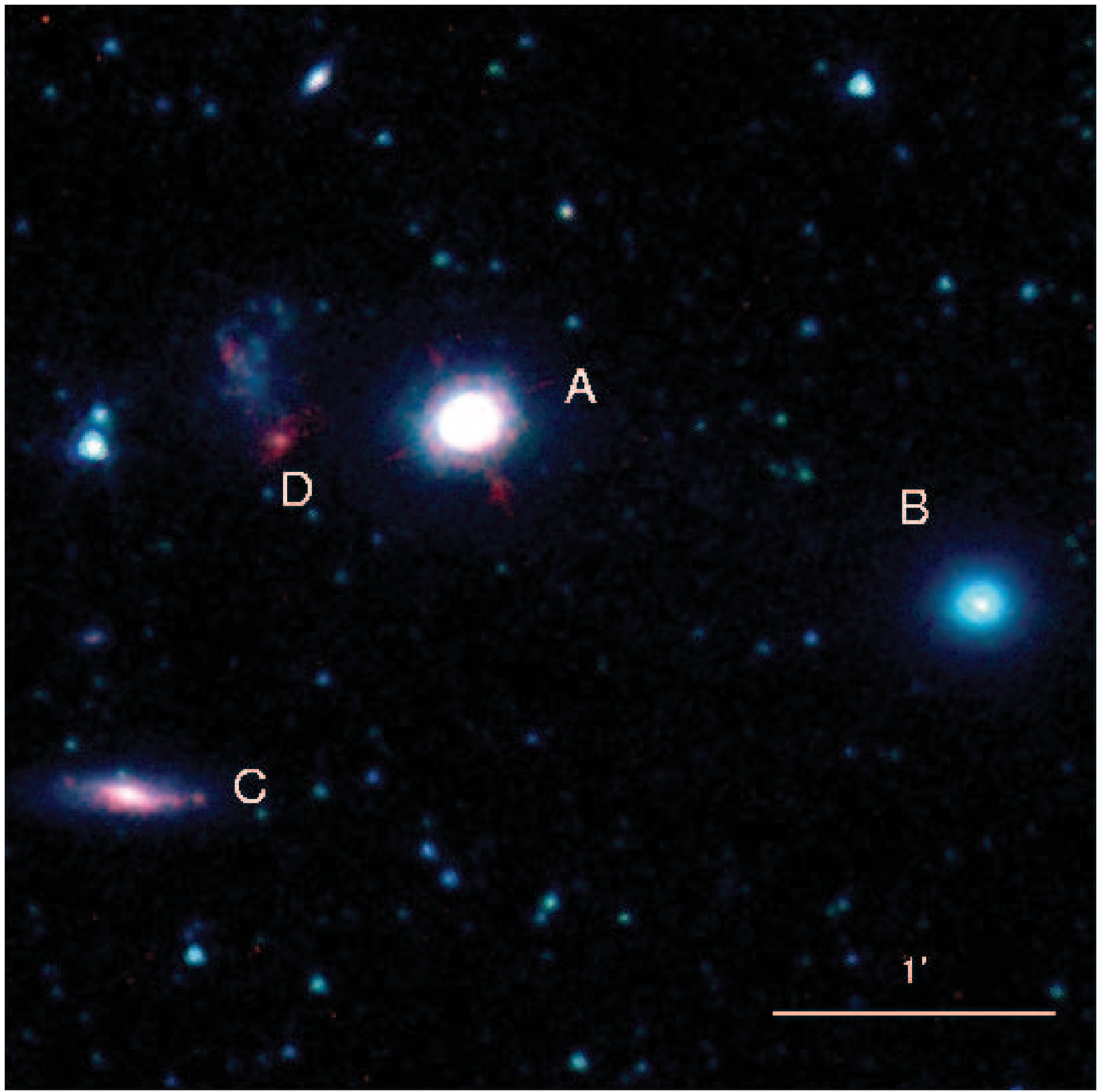}
	\caption{\spit~3.6--8~$\mu$m images of HCG~59. We present these
	individually to
	contrast the features evident in different bands and in a
	combined image to view the
	entire mid-IR picture. Galaxy~A is extremely bright in the
	hot dust/PAH-dominated
	5.8 and $8~\mu$m bands. Galaxy~B is faint in those bands,
	owing to little emission from
	heated dust. C is consistently bright in the four bands, given
	its ongoing star formation.
	D appears clumpy at long wavelengths, indicating patchy
	dust, heated by forming stars.
	In the 3.5 and 4.6\micron\ images, we also see faint, diffuse
	emission in the region
	between~A~and~B, which we study in
	Sections~\ref{sec:campanas}~and~\ref{sec:bplusr}. 
	}\label{fig:finder-spit}
	
%\end{center}
\end{figure*}

The optical imaging was complemented by \spit\ imaging in the
mid-infrared (IRAC 3.6--8~\micron\ and MIPS 24~\micron\ observations)
presented in J07 and shown in
Figure~\ref{fig:finder-spit}. In addition to the Rayleigh-Jeans tail
of stellar photospheric emission, the IRAC bands probe the presence of
hot dust and polycyclic aromatic hydrocarbons (PAHs), while the
$24~\mu$m observations trace cooler thermal dust emission.  The dust and PAH
emission are both stimulated by star formation activity. The harder
spectra of active galactic nuclei typically destroy PAH molecules
while heating dust to hotter temperatures than found in galaxies with star
formation alone.  At low AGN luminosities, the IR SEDs are often
ambiguous 
\citep[particularly in the presence of star formation; e.g.,][]{gallagher08}.

%%%--- Figure 5: Spitzer images ---%%%

\begin{figure}[htbp]
\begin{center}

	\includegraphics[width=1.1\linewidth, angle=0]{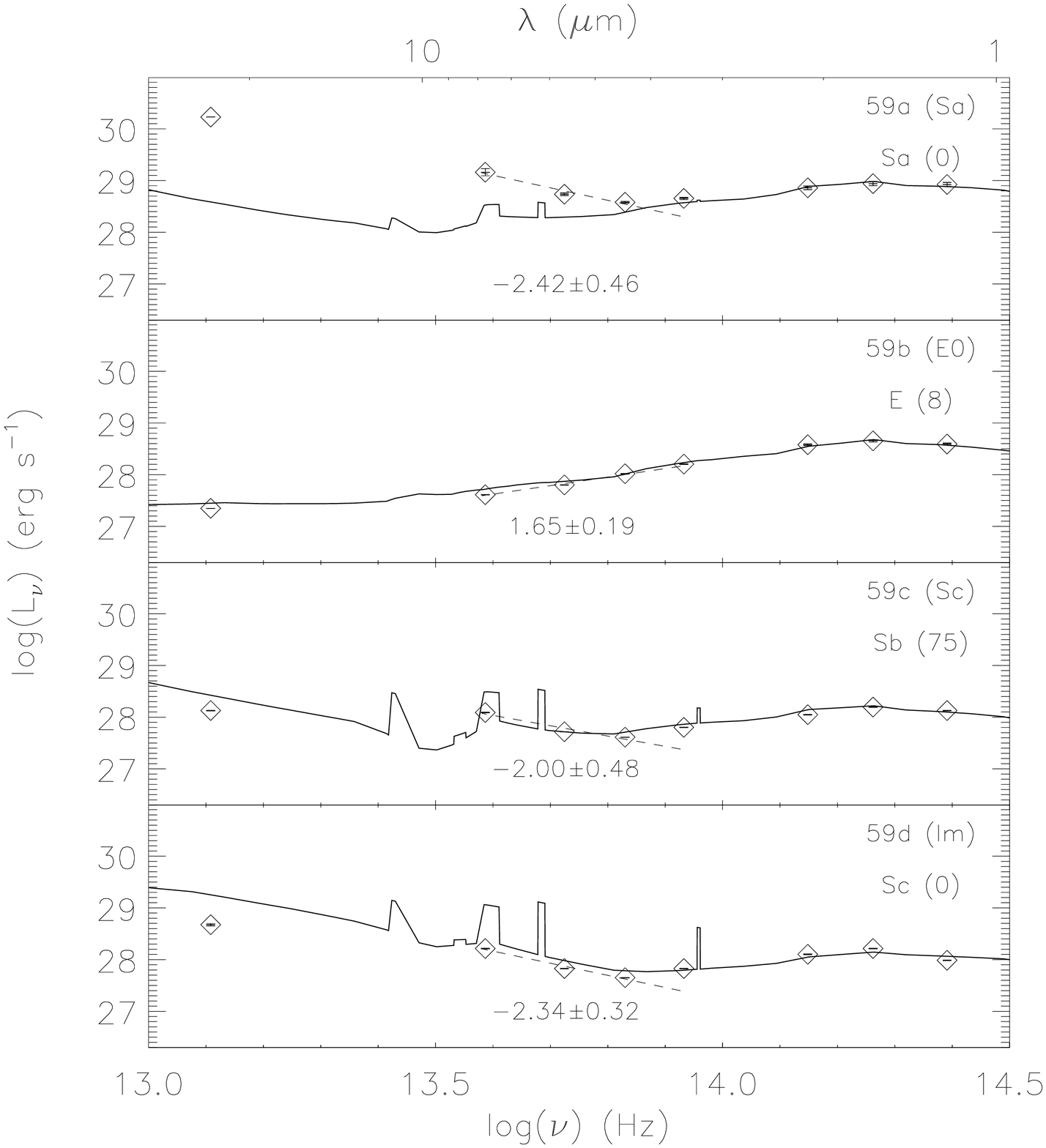}\\
	\caption{Near-to-mid IR SEDs for the primary HCG~59 members. The 
	photometric data, shown as diamonds, are drawn from \twomass\ 
	(\textit{JHK$_s$}) and \spit\ MIPS/IRAC~{3.5--24~$\mu$m} (from J07). 
	We annotate each panel with the morphological type from the literature 
	(top row), as well as the type chosen to most accurately represent 
	the data (bottom row). The bracketed number quotes the age in Gyr 
	of the template for the elliptical, or the inclination in degrees 
	of the spirals. We also cite the value of the \airac\ diagnostic
	\citep{gallagher08}, which is fit on the wavelength range indicated 
	by the dashed line. The solid curves represent GRASIL templates 
	\citep{silva98} appropriate for the morphological type of each galaxy 
	(see text). Galaxies~B~and~C are well represented by their nominal 
	templates, while A and D deviate: in the case of galaxy~D this is 
	due to its irregular nature, while we consider the nuclear activity 
	of A to be the source of the excess mid-IR emission 
	(see Section~\ref{sec:hcg59a} for details).
	}
	\label{fig:seds}	
\end{center}
\end{figure}

The \spit\ images were combined with \textit{JHK$_S$} observations
from \twomass~\citep{2mass} to plot the IR spectral energy
distribution (SED) of each galaxy (following J07),
presented in the frequency-space plot of Figure~\ref{fig:seds}. We
have used the \citet{silva98} templates for galaxies of various
morphological types. These map the SED of different galaxies as the
sum of starlight and gas and dust emission from star formation and
interstellar cirrus. We calculate the spectral index of the SED within
the IRAC bands through a simple power-law fit. This was defined by
\citet{gallagher08} as \airac, and it serves as a measure of star
formation activity. In logarithmic frequency units, the flux difference from 8
to 4.5$\,\mu$m leads to a positive gradient in quiescent environments,
while star formation registers as a negative slope. For HCGs, the
steepness of the slope is sensitive to the specific star formation
rate \citep{tzanavaris10}.

In brief, the SEDs of galaxies B~and~C follow their morphological
types of E/S0 and Sc, while the irregular nature of D does not allow for
a template to be assigned (we use Sc in
Figure~\ref{fig:seds}). Galaxy~A, nominally an Sa, shows strong excess
light at wavelengths associated with PAH and/or hot dust emission
($\lambda\geq5.8~\mu$m). If the flux in this region is associated with
star formation, an Sc template would be more appropriate. This is,
however, inconsistent with the optical morphology of A. Furthermore,
the shape of this emission is also consistent with the AGN dust bump
\citep[e.g.,][]{elvis94,gallagher08}, so in Section~\ref{sec:hcg59a}
we will examine the nuclear activity of this galaxy.

%%%--- Figure 6: IR SEDs ---%%%

The SED-fitting process could not be carried out for the new dwarf
galaxies catalogued in Section~\ref{sec:campanas} due to the limited
spatial coverage of the \spit\ imaging and the faintness of the dwarfs
(they were not detected in 2MASS).  As of July 2011, the Wide-field
Infrared Survey Explorer (WISE) photometric catalogue did not cover
HCG~59.

% Chandra -----------

\subsection{The high energy picture: \chan-ACIS observations}

\label{sec:xrays} 

\begin{figure*}[htbp]
	\begin{center}
	
	\includegraphics[width=\textwidth, angle=0]{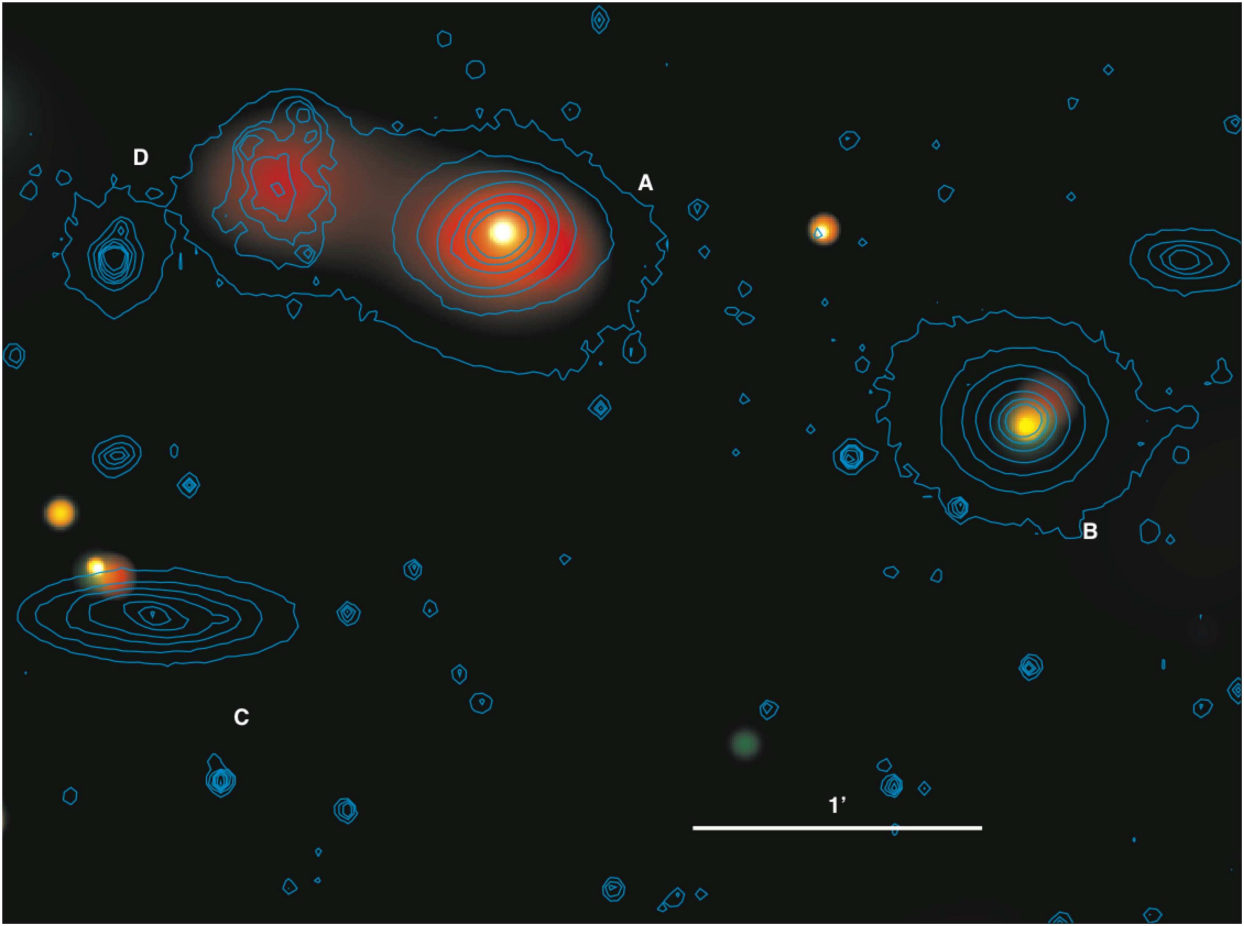}
	\caption{Adaptively smoothed three-band X-ray image of HCG~59,
	with overplotted \ib-band contours. Red, green and blue colors 
	represent soft~($0.5-2~$keV), medium~($1-4~$keV) and 
	hard~\mbox{(2--8~keV)} X-ray emission. Galaxy~C~(Sc) has no 
	detectable X-ray emission above the background level (the source
	at the limit of its optical isophotes is a likely background galaxy), 
	while D~(Im) shows some soft diffuse emission. The point source 
	detected in the nuclear region of galaxy~A~(Sa) is detected in 
	the full band with $L_{X ({\rm 0.5-8.0})} = 1.1\times10^{40}~$erg~s$^{-1}$.
	This is surrounded by diffuse emission. We resolve two point 
	sources in the central region of Galaxy~B~(E/S0) with 
	$L_{X ({\rm 0.5-8.0})} = (1.4, 1.7)\times10^{39}~$erg~s$^{-1}$.
	Section~\ref{sec:hcg59a} and Figure~\ref{fig:colmap} provide
	details. No diffuse emission is detected from the intra-group medium.
	}
	\label{fig:xray}
	
	\end{center}
\end{figure*}

The dataset is completed by \chan-ACIS data in the 0.5--8.0~keV
range.  Results reported here are drawn from the work by
Tzanavaris~\etal~(2011 in prep.).   
HCG~59 was observed by \chan\ between 2008-04-12 and 2008-04-13 at the
aim point of the back-illuminated S3 CCD of ACIS in very faint mode
with an exposure time of 39 ks (observation ID 9406, sequence number
800743, PI S. Gallagher). The data were processed using standard
\chan\ X-ray Center aspect solution and grade filtering, from which
the level 2 events file was produced.  Figure~\ref{fig:xray} shows an
adaptively smoothed 3-band X-ray image, with optical (\ib) contours
overplotted for comparison.

%%%--- Figure 7: X-ray map ---%%%

\texttt{Wavdetect}, the \texttt{CIAO 4.1.2}\footnote{http://cxc.harvard.edu/ciao} wavelet
detection tool \citep{freeman2002} was used in the
soft (0.5--2.0 keV), hard (2.0--8.0 keV) and full (0.5--8.0 keV) bands
to detect candidate point sources. The 1024$\times$1024
S3 chip field was searched with \texttt{wavdetect} at the $10^{-5}$
false-probability threshold. Wavelet
scales used were 1, 1.414, 2, 2.828, 4, 5.657 and 8.0 pixels.  Source
lists produced by \texttt{wavdetect} for each band were matched
against each other by means of custom-made scripts (K.~D. Kuntz,
priv. comm.)  to calculate a unique position for each candidate point
source, taking into account the varying size of the \chan\ PSF across
the S3 CCD.

Point source photometry was carried out for the objects in the source list using
\texttt{ACIS
Extract}\footnote{http://www.astro.psu.edu/xray/docs/TARA/AE.html} \citep{broos2010}
which takes into account the varying \chan\ PSF accross the CCD.
Poisson $\pm1\,\sigma$ errors on net counts were calculated by means
of the approximations of \citet{gehrels1986}.  Sources with measured
net counts smaller than the $2\,\sigma$ error were flagged as
non-detections.  Note that this method produces very similar results
to choosing a binomial probability threshold of 0.004 in \texttt{ACIS
Extract} \citep{xue}.

As sources have too few counts for reliable spectral fitting, we apply
the method of \citet{gallagher2005} to obtain a rough estimate of the
spectral shape by using hardness ratios, defined as HR $\equiv (H-S) /
(H+S)$, where $H$ and $S$ represent the counts in the hard and soft
bands, respectively.  Briefly, we use the X-ray spectral modeling tool
\texttt{XSPEC} \citep{arnaud1996}, version 12.5.0, to simulate the
instrumental response and transform observed HR values into an
effective power law index $\Gamma$ (where \mbox{$f_X \propto
E^{-\Gamma}~$photons~cm$^{-2}$~s$^{-1}$~keV$^{-1}$}), and also obtain
associated X-ray fluxes and luminosities.  This modeling includes
neutral absorption from the Galactic $N_{\rm H}$ of 
$2.6\times10^{20}$~cm$^{-2}$ \citep{nh_ref2}.

We estimate flux limits of $f_X \gtrsim 2.7 \times 10^{-16}$ \flux\
(0.5 - 2.0~keV) and $ f_X \gtrsim 1.6 \times 10^{-15}$~\flux\ (2.0 -
8.0~keV), corresponding to luminosity limits of
$L_X=1.3\times10^{38}$~\lumin\ (0.5 - 2.0 keV) and
$L_X=7.7\times10^{38}$~\lumin\ (2.0 - 8.0 keV).  Assuming these
limits, we use the $\log N - \log S$ relation of
\citet{cappelluti2007} to estimate the number of background sources we
would expect to detect in the HCG 59 field over the $1.96\times
10^{-2}$ square degree area of the S3 chip.  This number is $\sim 25$
and $\sim 20$ in the soft and hard bands, respectively, with a $\sim
20$\%\ uncertainty.  In the much smaller area ($3.66 \times 10^{-4}$
square degrees) covered by our galaxies, we expect
$< 1$ background source in each band.

We detect a total of 40 sources in the soft band and 33 sources in the
hard band over the ACIS S3 field.  We thus expect about 15 soft and 13
hard sources to be point sources associated with HCG 59.  We find that
9 soft and 3 hard sources are located inside the boundaries of the
MIR-based HCG 59 galaxy regions of J07.  We note that 11
sources that have only hard-band emission are located far from the HCG
59 galaxies and are thus likely background AGN.

In the central regions of two group galaxies, there are three notable
X-ray point sources detected with high significance.  As these point
sources are all with 5\arcmin\ of the \chan\ optical axis, the X-ray
positions are the ACIS Extract ``mean positions of events within the
extraction regions''
\footnote{See Section~5.3 of the ACIS Extract User Manual -- 
	http://www2.astro.psu.edu/xray/docs/TARA/ae\_users\_guide.html}
with intrinsic positional uncertainties of a few tenths of an
arcsecond.  There is some additional uncertainty from matching the
absolute reference frames of \chan\ and \hst, but this is expected to
be small as both are consistent with 2MASS at the $\sim0.1\arcsec$
level.

The first source, in galaxy~A, has a full-band luminosity of $L_{X
({\rm 0.5-8.0})} = 1.1\times10^{40}$~\lumin\ and an estimated $\Gamma
= 1.3\pm0.3$. The X-ray position of this source is 0.7\arcsec\ from
our quoted optical position (Table 1).  This isolated point source in
the nuclear region of galaxy~A has a luminosity that is consistent
with known low-luminosity AGN \citep[e.g.,][]{Ho2001} and
significantly higher than individual, luminous X-ray binaries.
The two point sources found in the central region of galaxy~B have
X-ray positions 0.2\arcsec\ and 1.3\arcsec\ from the \hst\
$i$-band galaxy centroid position (Table 1). Unfortunately, these
sources have fewer than 10 counts in each band, precluding even a
rough $\Gamma$ estimate.  Their full-band luminosities are $L_{X ({\rm
0.5-8.0})} = (1.7, 1.4)\times10^{39}$~\lumin, respectively.

The \chan\ images are also sensitive to diffuse emission from MK degree gas. 
As can be seen in our adaptively smoothed image, some soft, diffuse
emission is detected in galaxy D~(Im), likely associated with star
formation, as well as in galaxy A. In both cases, the diffuse emission
covers an area several times the size of the \chan\ PSF at that
location.  We obtain an upper limit on the IGM surface brightness as
follows.  We calculate the count rate in a source-free region between
group galaxies, and estimate the corresponding flux for $kT = 0.5$~keV
thermal emission using PIMMS.\footnote{http://heasarc.nasa.gov/Tools/w3pimms.html}  
We thus
estimate the IGM surface brightness to be $\lesssim 7.3 \times
10^{-17}$~\flux\ arcsec$^{-2}$.  Finally, we note
 that dwarf galaxy I is not coincident with any detected X-ray
 sources; none of the other new dwarf members of HCG~59 are within the
 \chan\ field of view.

 We will discuss the implications of these observations further in
 Section 4.

%
%=======  STAR CLUSTERS =======

\section{The young and old star cluster populations: star formation
over a Hubble time}\label{sec:clusters}

\subsection{Young star clusters and the past $\sim\,$Gyr of star
formation}
\label{sec:scs} 

The population of star clusters is representative of star formation as
a whole in any system \citep[\eg][]{bressert10}. They are formed {\it en
masse} after large events
\citep[e.g.,][]{gelys07b,isk08,isk09a,isk09b,bastian09antennae} and at a slower
pace at all times when a galaxy is forming stars
\citep{ladalada03}. The extreme brightness of young clusters makes
them detectable to large distances and therefore a reliable tracer
of the star formation history of their host galaxy over a 
Gyr or so. Beyond that point in time they are referred to as intermediate
age clusters and eventually globular clusters. As a whole, the cluster
population of a galaxy can reveal its star-forming history over a
Hubble time.

In this section, we analyze the cluster populations of the four main
galaxies in HCG~59 using the \hst\ PSF photometry described in
Section~\ref{sec:sc-phot}. We use color-color (CC) and color-magnitude
(CMD) diagrams to roughly age-date the clusters by comparing them to
evolutionary tracks. The \bvi\ filter combination of our \hst\ images
lacks coverage below $\sim4000~$\AA, which is crucial to breaking the
age-reddening degeneracy (owing to the inclusion of the Balmer jump
and near-UV continuum). However, it is still possible to infer the
passing of intense bursts of star formation via the clumping of
data-points along the evolutionary track, and unreddened young
clusters are clearly evident.

\begin{figure*}[htbp]
\begin{center}

	\includegraphics[width=\textwidth]{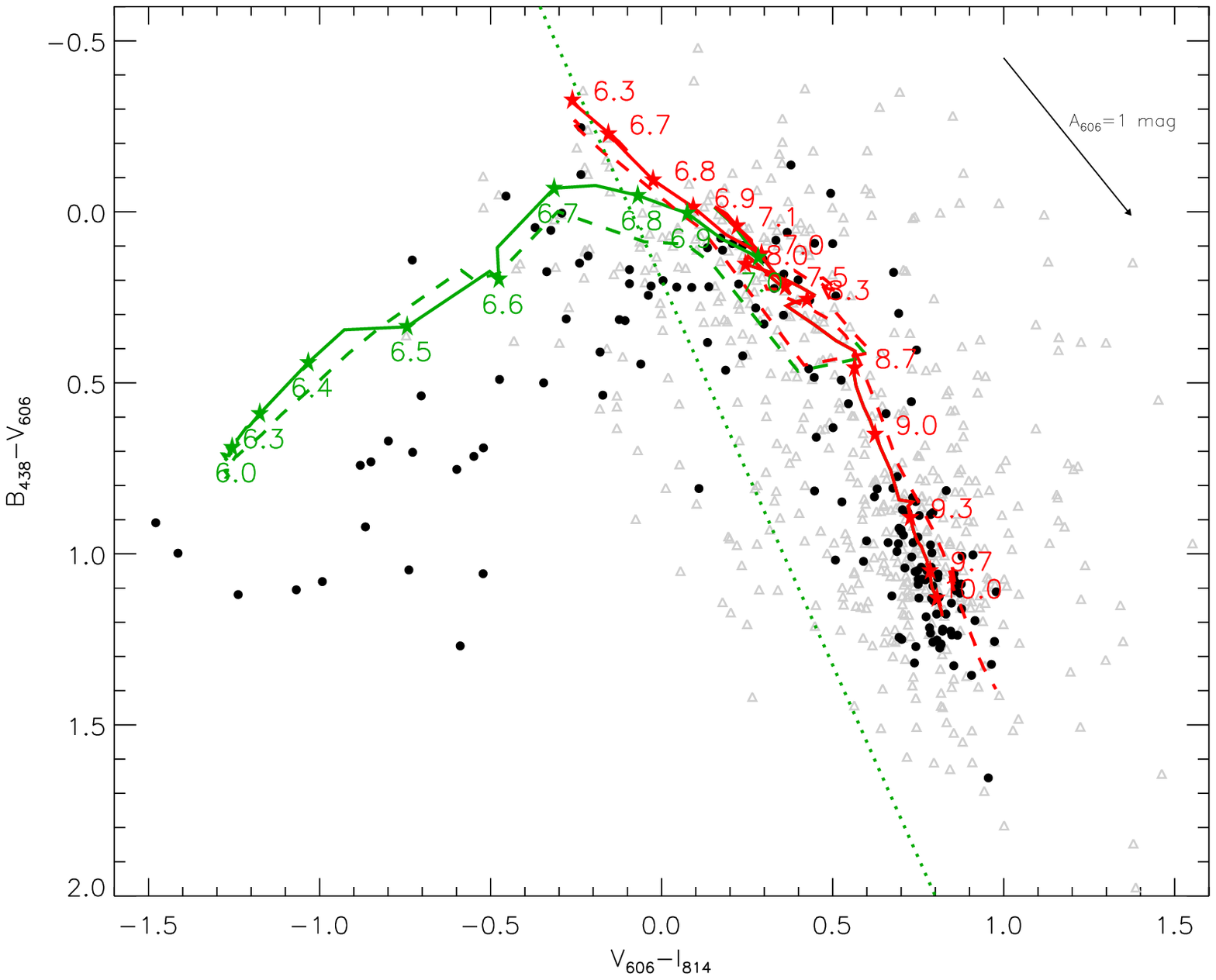}\\
	% HCG59_cc_all.eps
	\caption{\bb$-$\vb\ \textit{vs.} \vb$-$\ib\ colors of all SCCs in HCG~59, 
		plotted along with $\frac{1}{5}$~\Zsun\ (solid red line, bottom) and 
		\Zsun\ (dashed red) model tracks by \citet{marigo08}, spanning an age 
		range of 6~Myr to 13~Gyr (indicated by filled stars). The green lines 
		(solid and dashed for $\frac{1}{5}$~\Zsun\ and \Zsun) show a Starburst99 
		track that accounts for nebular emission, as suited to the very youngest 
		clusters that have not yet expelled their natal gas. Data-points that 
		lie to the left of the dotted green line are considered `nebular', \ie\ 
		younger than $\sim10~$Myr. The high-confidence sample of SCCs (those 
		with $M_V<-9$) is denoted by solid dots, while open triangles show 
		fainter candidates comprising the extended sample
		(all cluster candidates to the detection limit). Much
                of the spread in the
		extended sample colors is due to stochasticity in populating the stellar
		initial mass function within clusters, as discussed in the text. We 
		also indicate an extinction vector of length $A_{606}=1~$mag. We will  
		present the distribution of each galaxy separately in the next figure. 
	}\label{fig:colors-all}
	
\end{center}
\end{figure*}

\begin{figure*}[htbp]
\begin{center}

	\includegraphics[width=\textwidth]{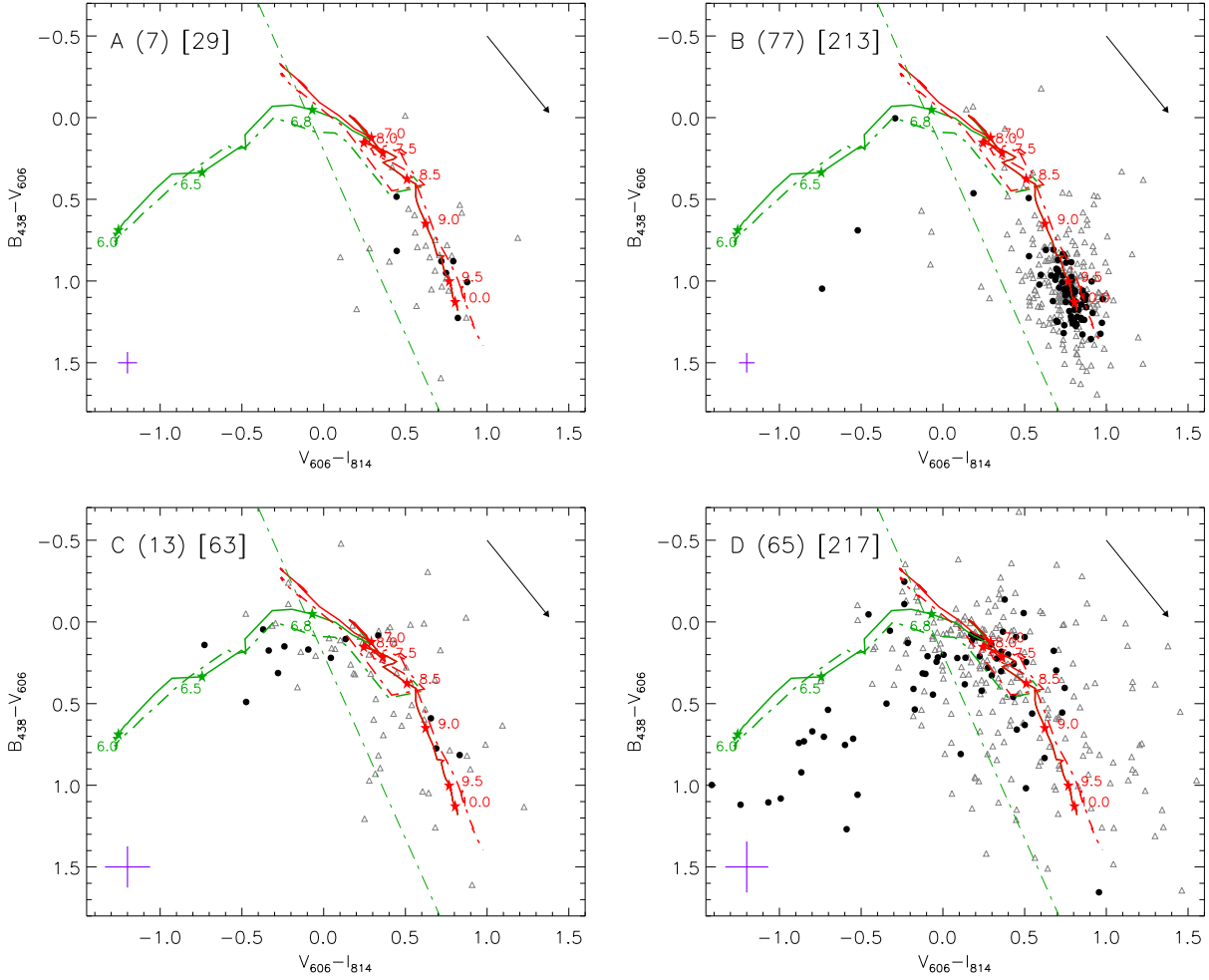}\\
	% HCG59_cc_regions.eps
	\caption{\bb$-$\vb\ \textit{vs.} \vb$-$\ib\ colors of SCCs, divided by 
	galaxy, following Figure~\ref{fig:colors-all}. Typical error bars are 
	indicated as crosses in the lower left part of each plot.  The irregular 
	galaxy~D is the only one that displays pronounced current star formation
	activity, as deduced by the number of nebular sources -- although more 
	nebular sources could be hidden behind dust and gas in the highly inclined 
	spiral galaxy~C. Galaxies~C~and~D are consistent with continuous star 
	formation histories over the course of their histories. The absence of 
	GCs in D might indicate the galaxy is young, although it could be due to 
	the fading of its low-mass cluster population below our detection limit 
	(which is a function of age). Galaxies~A~and~B only appear to host globular 
	clusters and a handful of intermediate-age and young clusters, indicating 
	quiet star formation histories over the past Gyr or so. 
	}\label{fig:colors}
	
\end{center}
\end{figure*}

\begin{figure*}[htbp]
\begin{center}
	
	\includegraphics[width=\textwidth]{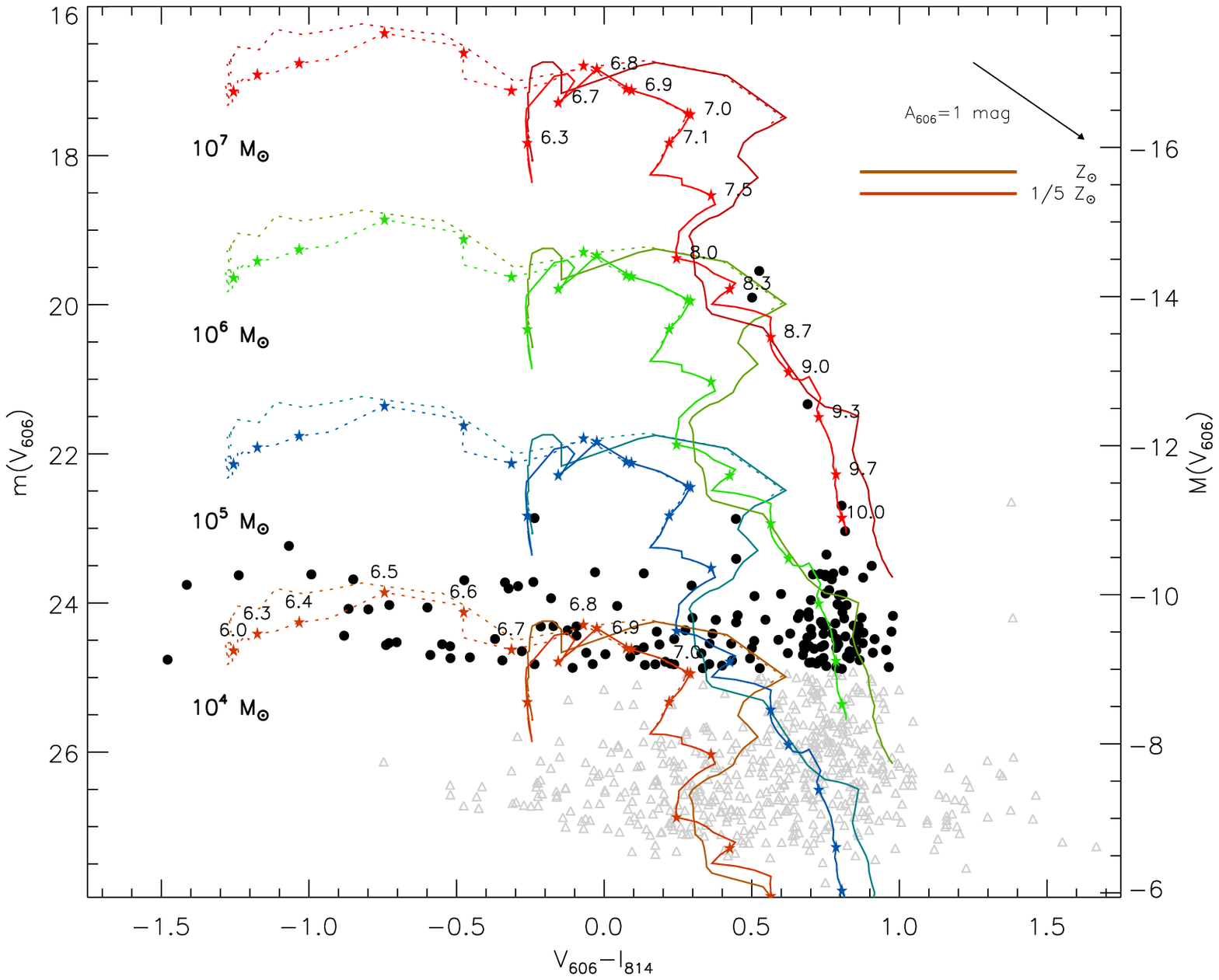}\\
	% HCG59_cmd_all.eps
	\caption{\vb\ \textit{vs.} \vb$-$\ib\ color-magnitude diagrams, 
		following the datapoint plotting conventions of 
		Figure~\ref{fig:colors-all}. We have provided \citet{marigo08} model 
		tracks for four different cluster masses, times two metallicities, 
		$\frac{1}{5}$~\Zsun\ and \Zsun\ (top and bottom) (solid lines), with 
		the `nebular' tracks plotted as dotted lines. As in 
		Figure~\ref{fig:colors}, the age increases from top left to bottom 
		right. In both samples, we tentatively deduce typical cluster masses 
		-- with the caveat that mass and extinction are degenerate --  of 
		$\lesssim10^4$~\Msun, with an additional locus of possible 
		$10^6~$\Msun\ clusters. This distribution is consistent with 
		local cluster populations, given the imposed cut-off for the 
		high-confidence sample at $M_V<-9$ (and therefore the lower limit of 
		$\sim10^4~$\Msun). 
	}\label{fig:cmd-all}
	
\end{center}
\end{figure*}

\begin{figure*}[htbp]
\begin{center}
	
	\includegraphics[width=\textwidth]{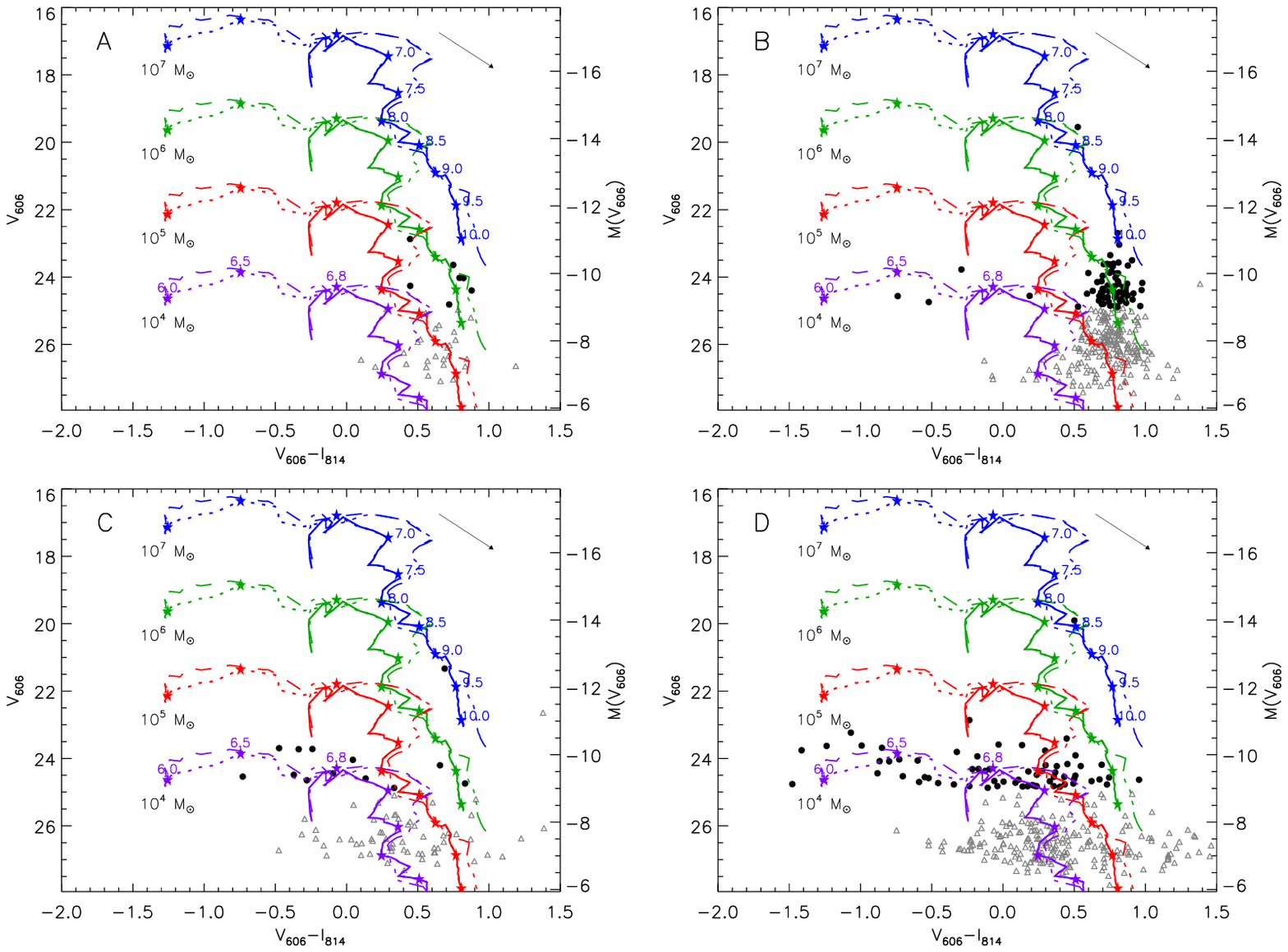}\\
	% HCG59_cmd_regions.eps
	\caption{\vb\ \textit{vs.} \vb$-$\ib\ color-magnitude diagrams
	for each galaxy individually,
		following the plotting conventions of
                Figure~\ref{fig:cmd-all}.  Galaxy B is notable for the
                large population of apparently old, massive
                ($\sim10^6$~\Msun) clusters,
                while galaxy D has shows a sizable population of
                $10^4~$\Msun\ clusters consistent with a
                predominantly young and intermediate-aged system.    
		}\label{fig:cmd}

\end{center}
\end{figure*}

Figures~\ref{fig:colors-all}~through~\ref{fig:cmd} show the
high-confidence sample (\ie\ clusters with $M_V<-9$~mag; see
Section~\ref{sec:sc-phot}) as solid dots, while the extended sample
is marked with open triangles.  The solid and dashed red lines
(running top to bottom) show \citet{marigo08} evolutionary tracks for
simple stellar populations (SSP) of $\slantfrac{1}{5}$~\Zsun\ and
\Zsun. This is slightly different from our previous work, where we
used \citet{bc03} models. We made this change because the
\citet{marigo08} models seem to provide a good fit to both young
clusters and globulars, unlike other model suites which
focus on one part of the cluster population. The green lines that run
more or less horizontally show Starburst99 \citep[SB99;][]{sb99} tracks of the same
metallicities. These also include nebular emission, which we expect to
often be present during the first $\sim10$~Myr of evolution. At this
state the cluster is still surrounded by residual gas from the time of
its formation ionized by UV photons. This short-lived phase ends when
the first stars evolve and explode in supernovae that expel the
gas. Clusters with colors redder than 0.8 in both axes are most likely
GCs, although they might be highly reddened young clusters.

When contrasting the extended and high-confidence samples, the former
appears to spread more in color-space. This reflects a mass effect
intrinsic to star clusters, rather than indicating contamination. Two
recent studies, \citet{silvavilla11} and \citet{popescu10},
independently reached the conclusion that lower mass (fainter)
clusters, $M\lesssim10^4~$\Msun, often exhibit deviations from the
theoretical model tracks. In this mass regime, the underlying stellar
initial mass functions (IMF) are under-sampled and stochastic effects
dominate the overall light. Given that the IMF is populated randomly,
it is physically equivalent to creating either one high mass star, or
$\sim100$ low mass stars (given the IMF slope).  In contrast, a
high-mass cluster will populate the IMF fully.  Consequently, a
population of high-mass clusters will have smaller intrinsic
photometric dispersion.  A cluster with lower mass will run out of
material before the IMF becomes fully sampled, thus leading to the
presence of gaps and spikes. Because of this effect, the presence of a
high mass star in a low mass cluster will make the cluster appear to
have a larger photometrically derived mass than one with only lower
mass stars.

We now treat the population of each galaxy individually.  In the
following paragraphs, all star and globular cluster candidates
discussed are taken from the high-confidence sample.  The distinctions
of young, intermediate, and globular are inferred from the locations
of candidates along evolutionary tracks within the \bb--\vb\
vs. \vb--\ib\ color-color space.

{\bf Galaxy~A} (Sa) hosts a very small detectable population of five
GCCs and two intermediate-age cluster candidates. The CMD shows them
all to be consistent with having high masses, with \mbox{$\log(M/\Msun)>5$}. 
The lack of young clusters, combined with the high masses
of the ones detected, indicates that galaxy~A has stopped forming
stars at a high rate.  Because the mass-to-light ratio of a stellar
population increases with age (\ie, the older a cluster becomes, the
higher its mass must be for detection), young clusters should
dominate the cluster population when the SFR is high.  The
non-detection of {\em any} young clusters is further evidence that the
UV+IR emission is dominated by AGN continuum, and the inferred SFR of
5~\msun~yr$^{-1}$ is severely overestimated.

{\bf Galaxy~B} (E/S0) hosts mostly red clusters; the GCCs vastly
outnumber young cluster candidates in this system. In fact, the four
SCCs in the high-confidence sample do not strictly belong to galaxy~B, but are
found in the IGM: two nebular sources are located in a clump to the
west of the galaxy, which we are treating as part of a stream that
connects this galaxy to the dwarf HCG~59I in our field of view.  We
will return to these H\two\ regions in Section~\ref{sec:bplusr}. The
two non-nebular SCCs are distinct clusters in dwarf galaxy~I.  This
means that the presence of patchy dust (as seen in the \hst\ images of
59B) by itself is not indicative of star formation at a high enough
level to produce massive clusters, consistent with its SFR of
0.02~\msun~yr$^{-1}$.

{\bf Galaxy~C} (Sc) shows a continuous star formation history, evident
in the smooth distribution of datapoints along the evolutionary
track. The presence of nebular sources indicates some current star formation,
while the low masses ($\approx10^{4}$~\msun) derived throughout the
sample imply an overall low level of star formation over time. This is in accord
with the low value of 0.16~\msun~yr$^{-1}$ for the SFR of the
galaxy. There is no pronounced GC population.

{\bf Galaxy~D} (Im) is an unusually large irregular galaxy.  It
shows a continuous star formation history
through to the present. The CMD shows a handful of SCCs with 
\mbox{$\log(M/\Msun)\gtrsim5$}.  At 0.48~\msun~yr${^-1}$, 
this galaxy has the largest SFR
of those with young clusters.  There is no old component in the
cluster distribution, no evident halo of GCCs, again possibly due to
the low mass of the system which implies a small globular
cluster system.
Furthermore, the extended sample does not reveal a tight, correlated color
distribution characteristic of GCs. The youngest clusters lie at a
typical color-space distance of $\sim0.3$~mag away from the nebular
model track along the reddening vector, indicating the presence of
dusty star-forming regions.

%%%--- Figure 8: Clusters, CC, all ---%%%

%%%--- Figure 9: Clusters, CC, panel ---%%%

%%--- Figure 10: Cluster CMDs, all ---%%%

%%--- Figure 11: Cluster CMDs, panels ---%%%

% Complexes
\subsection{Star cluster complexes}\label{sec:complexes}

One step above star clusters in the star formation hierarchy is
cluster complexes, large agglomerates of young stars, arranged in a
fractal distribution that follows the collapse of the progenitor
gas. These structures can be used to understand the global star
formation activity in a galaxy. They are found to be more compact at
higher redshifts than in the local universe~\citep[with two to five times
higher mass surface density, as found by][]{elmegreen09}, although in
one local interacting compact group, HCG~31, we find complexes to be
similar to those at intermediate redshifts~\citep{gallagher10}. We
argue in this series of papers that compact groups might process gas
more efficiently when interacting than most other environments apart
from the infall regions of galaxy clusters \citep{walker10}. HCG~31
fits well in that context.

\begin{figure*}[htbp]
\begin{center}
	
	\includegraphics[width=\textwidth]{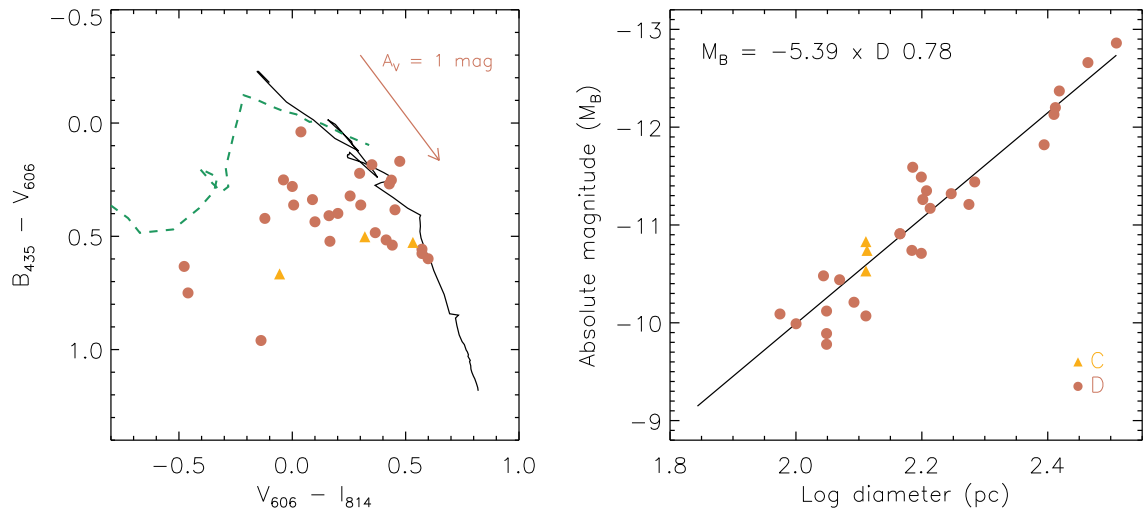}\\
	% complexes.eps
	\caption{Color-color and size-luminosity ($D-M_B$) diagrams for star 
		cluster complexes in HCG~59. The plotting of the
		evolutionary tracks follows the conventions of
		Figure~\ref{fig:colors-all}.  Complex symbols are
                coded to indicate where they are found: galaxy C
                (filled, yellow triangles) or D (filled, orange circles) 
		The colors of the complexes are indicative of star formation, as 
		expected. We find the complexes to follow a linear luminosity-size 
		relation (see the upper left corner of the right-hand panel) very well. This implies a
		similar surface brightness for
		complexes across the group, unlike in HCG~7, where this relation is 
		not obeyed.
		}
	\label{fig:complexes}
	
\end{center}
\end{figure*}

Star-forming complexes are extended amorphous regions, with dust
scattered across their surfaces. Their boundaries were identified by
eye and measured by contours down to a limiting surface brightness
about 10$\sigma$ above the background.  We single out 30 star-forming
complexes in this compact group, three in galaxy~C, and 27~in~D.  We
find no complexes in the E/S0 galaxy~B~or~A, the evolved (Sa) spiral. The
sizes of these complexes are comparable to those in local systems
\citep[Figure~\ref{fig:complexes}; \cf][]{elmegreen94, elmegreen96}
and consistent with HCG~7, a relatively inactive compact
group~\citep{isk10}. Their colors are indicative of star formation and
nebular emission, as expected of star-forming regions
(Figure~\ref{fig:complexes}, right). In all, the star-forming
complexes in HCG~59 are comparable to their counterparts in local
star-forming galaxies. They follow very closely a relation between
brightness and size, much more so than we found in our previous study of HCG~7.
We show this correlation in Figure~\ref{fig:complexes} (right) where
the line is a simple linear fit to the data. The linear fit implies a
similar surface brightness for complexes across the group, unlike
in HCG~7, where we found significantly more scatter.

%%--- Figure 12: Complexes ---%%%

%======= GLOBULAR CLUSTERS =======
\subsection{The ancient globular cluster systems of HCG~59}\label{sec:gcs}
Globular clusters represent the earliest eras of star formation in a
given galaxy.  Their color distributions carry the imprints of
interactions and mergers and thus may probe the history of their hosts
over a very long timescale.

\begin{figure*}[htbp]
\begin{center}
		
	\includegraphics[width=0.33\textwidth]{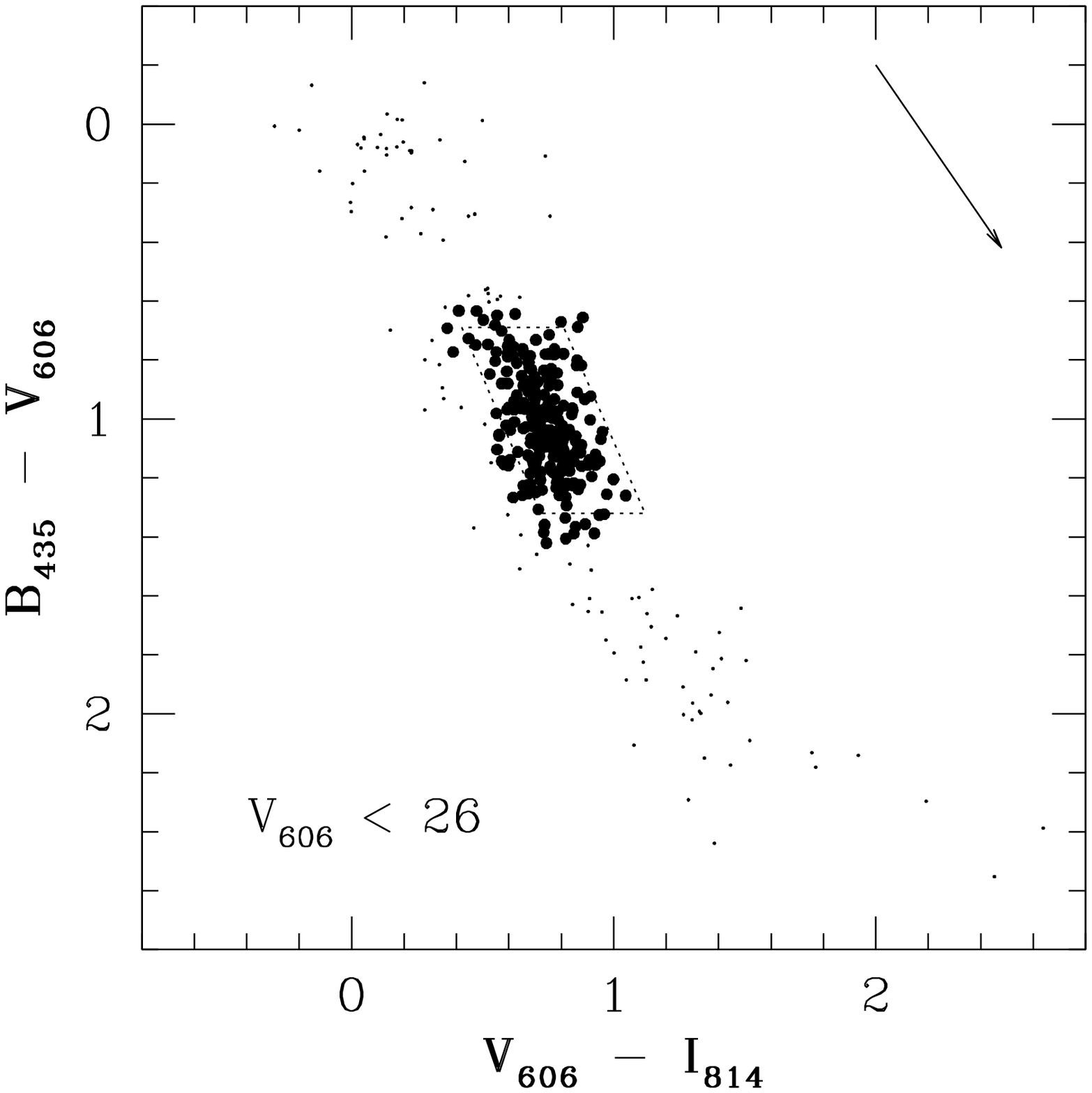}
	\includegraphics[width=0.31\textwidth]{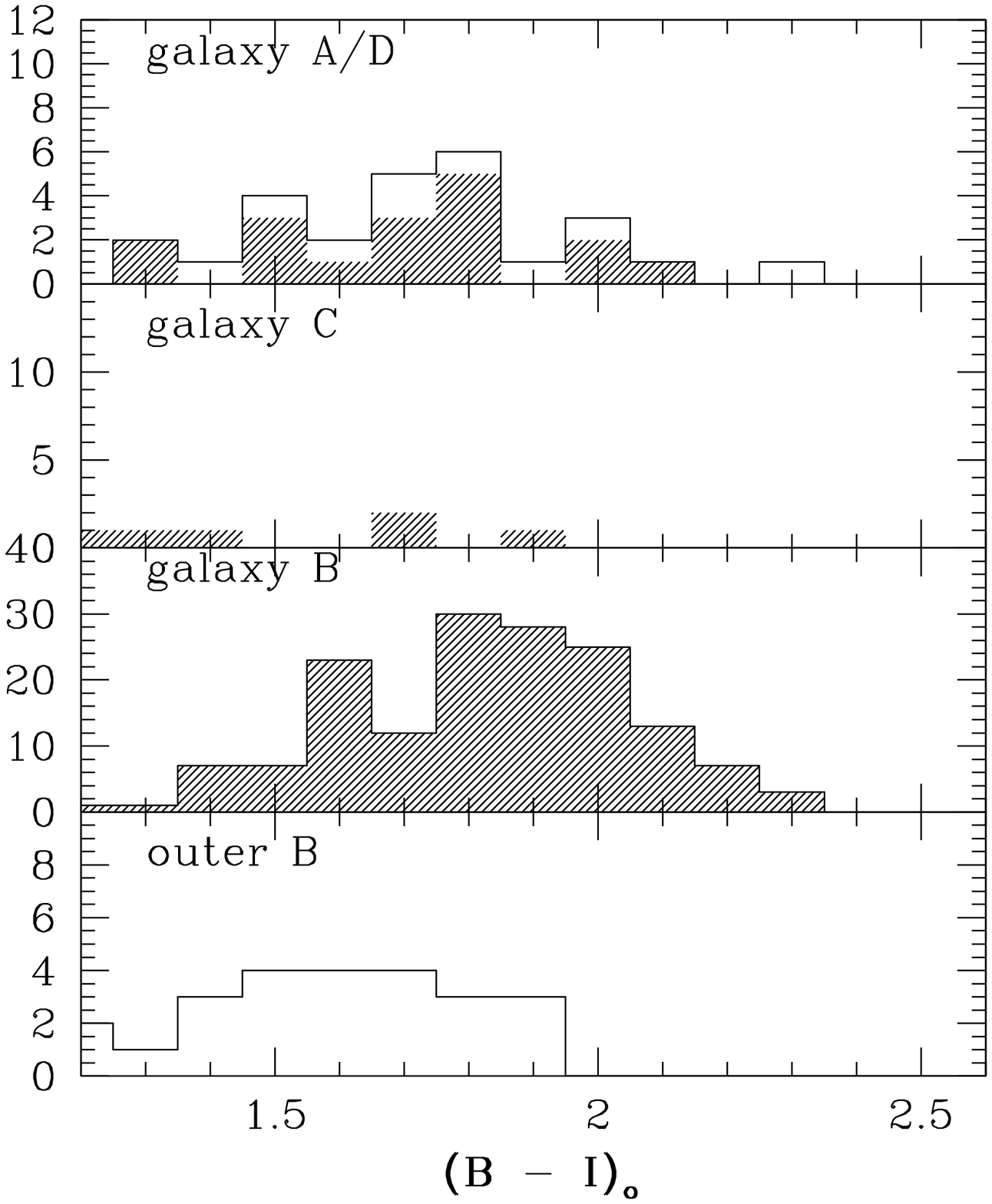}
	\includegraphics[width=0.33\textwidth]{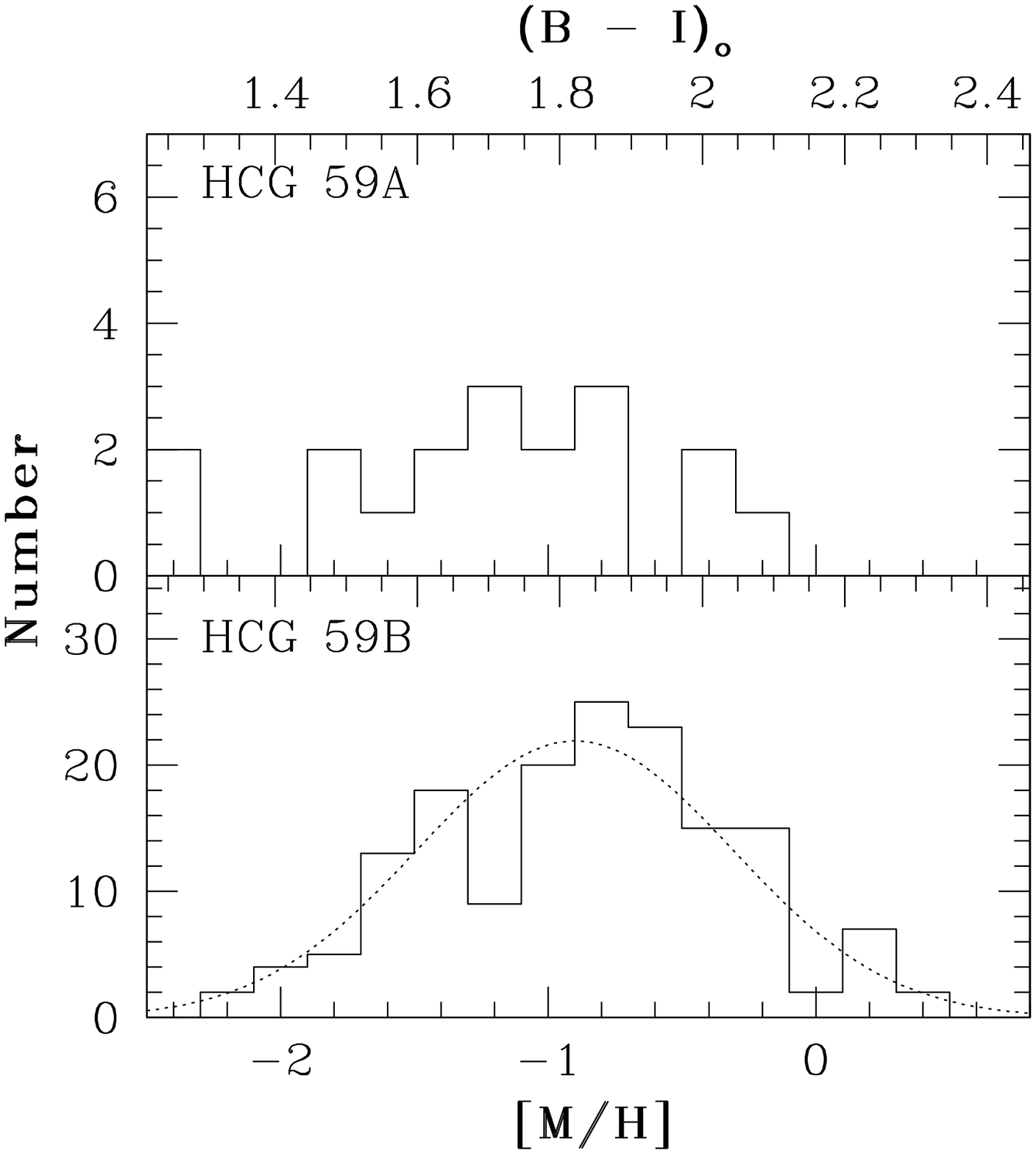}
	% f12_1.eps, f12_2.eps, f12_3.eps
	\caption{
		{\bf Left:}~Color-color diagram of all detected GCCs. The dotted 
		parallelogram on the left shows the selection box, based on the 
		colors of GCs in the Milky Way \citep{harris96}. 
		\mbox{{\bf Center:}~$(B-I)_0$ distribution} for all detected GCCs. 
		The shaded 
		area in the plot of the A/D distribution omits GCCs superposed 
		over the body of D. The `outer B' distribution covers all GCCs 
		in pointing~2 that are at least 15~kpc from the centre of galaxy~B. 
		This region might suffer from some contamination, although it is 
		consistent with observed distributions of GCs, which show a tendency for  
		red clusters to be at large galactocentric distance as
		would be expected for a halo population. On
		the {\bf right}, we relate the color distribution to
		that
		of metallicity and compare galaxies~A~and~B directly. The dotted line 
		in the GC metallicity distribution of galaxy~B shows a best fit 
		Gaussian, which we plot since a more complex distribution is not 
		statistically justified. 
	}\label{fig:gcs}
		
\end{center}
\end{figure*}

In this section we provide a full analysis of the number and color
distributions of the GC populations in HCG~59 and complement the star
formation and interaction histories we began to explore in
Section~\ref{sec:scs}.  This analysis is built around the plots of
Figure~\ref{fig:gcs}. The top row has a color-color plot (left) and histogram
of the $(B-I)_0$ distributions (right) in each galaxy. (The color-magnitude diagram is shown in the left panel Figure~\ref{fig:gc-diag}.) Owing to
their projected proximity, the GC systems of galaxies~A~and~D overlap, and so we
cannot provide separate analyses. However, given the relative masses
of the two galaxies and their evolutionary stages, it is safe to
assume that the population is dominated by A.

%%--- Figure 13: GC CCs ---%%%

We find a fairly large population of globular clusters across this
compact group, the vast majority of which are found in and around
B. Recall that given the fading of clusters with age, GCs need
extremely high masses, $M\gtrsim10^5~$\Msun, to be detected to such
large distances.
We also find some clusters at large radial distances from the center
of this galaxy, including some found along the stellar stream that
appears to connect galaxies~A~and~B (see Section~\ref{sec:bplusr}) and
its projection on the far side of galaxy~B.  Galaxies~A~and~C host
small populations. To quantify these populations, we derive the specific frequency, $S_N$,
a measurement of the number of clusters per unit galaxy luminosity,
for each galaxy.
First, we correct the observed number of GC candidates by
the background correction noted above, and then calculate the total
number of GCs expected around each galaxy by first adopting a
photometric completeness fraction of $f=0.9\pm 0.1$ for objects with
$V_{606}<26$, and correcting for the expected fraction of GCs that lie
below this magnitude limit. The $S_N$ values are 0.3, 7.7 and 0.1 for
galaxies~A,~B~and~C respectively, assuming that all bright GCCs
consistent with the halo of A are actually bound to A. The measured and
derived numbers are collected in Table~\ref{tab-gc}.

The size of the population in galaxy~C is consistent with 
its Sc morphological type, while B has a tremendously rich 
system, about twice the number of GCCs expected. Conversely, 
the GC population of galaxy~A is much poorer than expected, 
compared to the values of $S_N\sim 1$ typically seen in Sa 
spirals \citep{chandar2004}. Many of the detected 
GCCs appear to lie in a ring just outside of the bulge of the galaxy.
{Comparing to other GC populations in HCGs,  our specific
  frequency for HCG~59B of $S_N=7.7\pm 3.0$ is larger than (yet still
  consistent within the uncertainties) those observed so far in other
  large elliptical galaxies in HCGs: global values of $S_N=3.6 \pm
  1.8$ for HCG 22A from \citet{darocha02},\footnote{The lower $S_N\sim
  1.7\pm 1.0$ from \citet{barkhouse01} for HGC~22~A is due to their
  larger $M_V$ for the galaxy -- both studies find a very similar 
  total number of globular clusters.} and $S_N=4.4\pm 1.3$ for  
  HCG~90C from \citet{barkhouse01}.}

Inspection of the region marked `outer B' in Figure 3 shows a possible
excess of GC candidates over the background level described above in
Section 2.4, perhaps tracing an intra-group stellar population.
{This is not unusual, as compact groups by their very nature are likely to promote interactions. \citep{white03} detected diffuse intragroup emission in HCG~90 accounting for up to half the light of the group. Two further studies, \citet{darocha05} and \citet{darocha08}, found intra-group light in another six groups: HCGs~15, 35, 51, 79, 88, and 95. 
}
In HCG~59, a total of $N=15\pm 4$ GCCs lie within the 2.7 arcmin$^2$
region, where we expect a background contribution of $6\pm 2$
objects. Assuming only Poisson noise, this suggests a $\sim 2\sigma$
excess of objects in this part of HCG 59.  This possible excess could
be due to variations in the stellar density in the Galactic halo or
the Sgr Stream. However, the luminosity function of this small number of
sources in the `outer B' region (shown as the solid line in the right
panel of Fig. 3)
is weighted towards the faint end, suggesting that 
at least some of these objects are indeed true GCCs located far ($\sim
25-50$ kpc) from galaxy B.  The unusually large population of GCCs
associated with B, the anomalously small population in A, and the
possible population of GCs in the IGM (presumably stripped from a
member galaxy), all indicate that this compact group environment may
have redistributed GCs between member galaxies and/or to the IGM.

%SCG: to here

{In particular, possible interactions in the recent history of galaxy~B might
have redistributed its GC system. With this in mind, we compare the 
azimuthally averaged radial profile of the spatial distribution of 
the GC system with the surface brightness profile of the galaxy, as 
derived through GALFIT \citep{galfit}. The galaxy is best fit by a 
single-component S\'{e}rsic profile of shape $n=3.1$. Interestingly, we 
find the S\'{e}rsic profile of the galaxy to provide a better description 
of the GC system than the best fitting power-law profile of index $-1.3$. 
This is contrary to the finding of power-law GC system distributions 
around loose group member NGC~6868 and HCG~22~A. If that is to be 
considered the norm for compact groups, then perhaps the recent 
interaction activity about galaxy~B has changed the shape of its 
GC halo. }

%%%--- TABLE: GC population ---%%%
\input{tab4}
%%%

The bottom panel of Figure~\ref{fig:gcs} shows the $(B-I)_0$ color
distribution (converted to the Johnson photometric system for direct
comparison to Galactic globulars) of the clusters in each galaxy. The
shaded area in the plot of A/D shows clusters that are clearly part of
A, \ie\ omits the ones that are projected upon the body of D. This
does not alter the distribution, strengthening our assumption of a
small population in~D.  Galaxy~C also hosts a very small
population, as expected due to its low mass.  

The color distribution can act as a proxy of metallicity for GCs and
we take advantage of that to compare the two populous distributions of
galaxies~A~and~B. The color distribution of galaxy~A seems fairly flat
and the low numbers do not allow for a statistical
treatment. Galaxy~B, however, provides a large enough population to
perform a test for bimodality, using the KMM~algorithm of
\citet{ashman94}.   This returns no evidence for a composite distribution
in the metallicity distribution.

%=======  RESULTS =======
\section{Discussion}\label{sec:results}
%
% dwarf galaxies
\subsection{Extending the membership of HCG~59}\label{sec:dwarfs}
%
% ** ** ** WITH G ** ** **
%------
%[RV - <RV_giant>] / Sigma = 
%     0.164143   0.00796813      2.67889    0.0780877     0.298008
% 
%Dynamical mass:
%  of giants only =    2.9400441e+12
%  of all members =    2.7811880e+13
% 
%log(M_HI) / log(Mvirial): 
%  of giants only =       0.76112692
%  of all members =       0.70587899
%------
%
% ** ** ** WITHOUT G ** ** **
%------
%[RV - <RV_giant>] / Sigma = 
%     0.164143   0.00796813    0.0780877     0.298008
% 
%Dynamical mass:
%  of giants only =    2.9400441e+12
%  of all members =    1.0718727e+13
% 
%log(M_HI) / log(Mvirial): 
%  of giants only =       0.76112692
%  of all members =       0.72831124
%------
%
In Section~\ref{sec:campanas} we introduced a search for dwarf galaxies in 
HCG~59, which we continue here. Since all objects we are considering here are 
covered by SDSS, we will not provide images and spectra here.  More information can be obtained from the SDSS database using the plate IDs, Modified Julian Dates (MJDs), and fiber IDs given in Table~\ref{tab:dwarfs}.

Regarding the morphologies of the new members, I, the candidate
covered by our \hst\ imaging, seems irregular, with a peaked light
profile. This agrees with its spectrum, which shows clear emission
lines and a continuum shape typical of a spiral. We classify it as
dIm. The rest of the galaxies are not covered with high-resolution
imaging, so we are more conservative in classifying them. F shows
quite clear spiral structure in the SDSS images, and is therefore
given an Sd type. We note that its star formation seems to be
declining, given that H$\alpha$ is the only detectable %found purely in
emission line. G~and~H appear quite irregular and elongated, with spectra
exhibiting blended emission and absorption. We assign them dIr
types. Finally, J shows a morphology closer to spherical and weak
emission lines (although the S/N does not allow for a confident
determination); we assign it a dE type.

\begin{figure*}[htbp]
\begin{center}

	\includegraphics[width=\textwidth, angle=0]{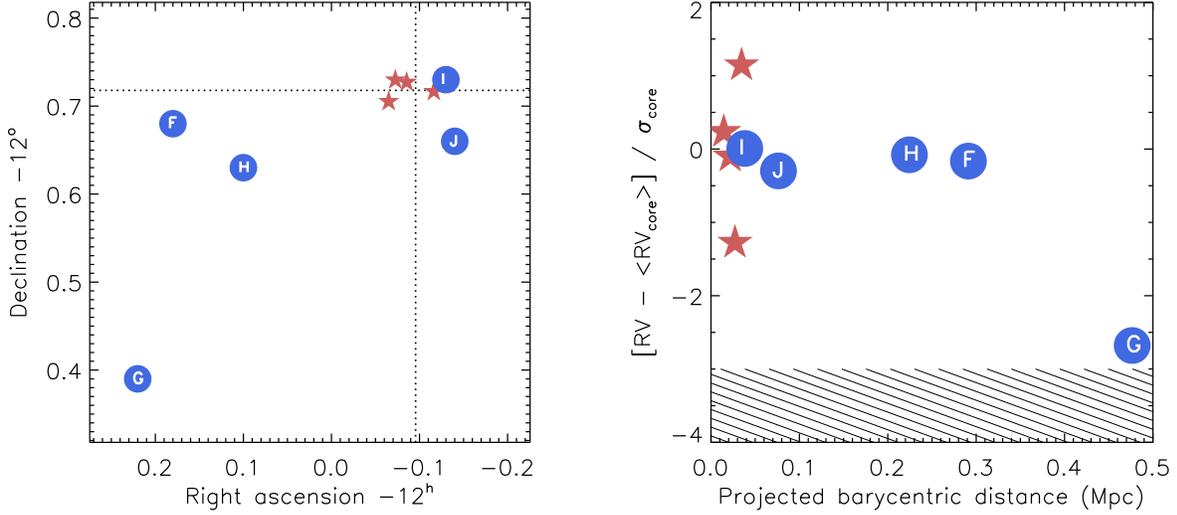}\\
	% pspace.eps
	\caption{Diagnostic diagrams of dwarf galaxy membership. The {\bf left} 
		panel shows the positions of the dwarfs and giant galaxies as blue 
		dots and red stars respectively. We have labelled the dots with the 
		appropriate lettering for HCG members.  
		On the {\bf right} we present a phase-space diagram that plots the 
		offset of a galaxy's radial velocity from the established group 
		mean ($\sigma_\textup{\scriptsize core}$, the line of sight velocity
		dispersion of the giant galaxies) against its distance from the group 
		barycenter (mass-weighted centroid). Four of five spectroscopically 
		detected dwarfs lie within the boundaries set by the main galaxies 
		and as such are confirmed as members. The fifth is  moving at a 
		relative radial velocity less than three times the group's velocity 
		dispersion and is therefore also included as a member. The lower 
		boundary of this $3\,\sigma$-clipping criterion is shown on the 
		plot as the shaded area. 
	}\label{fig:pspace}
	
\end{center}
\end{figure*}

Regarding their roles as group members, four of the five galaxies do
not appear to be interacting with any other members, as might be
expected from their locations far from the group core
(Figure~\ref{fig:pspace}, left).  It is only I that shows some
evidence of an interaction with galaxy~B, in the form of the `B-I arc'
described in the Section~\ref{sec:campanas}. This stellar stream includes 
the large star-forming region we find to the west of galaxy~B, which 
is part of the analysis of Section~\ref{sec:scs}.

In order to assess whether these galaxies belong to the group, we
perform a phase-space analysis, following the statistical studies of
\citet{mz98} and \citet{zm00}. This is shown in
Figure~\ref{fig:pspace}, right panel, where each datapoint represents
a dwarf galaxy. The x-axis measures the distance of a galaxy from the
group centroid, calculated as the mass-weighted average position of
the four main members (A through D). The y-axis shows the offset of a
galaxy's radial velocity from the group mean, normalized to the core
group (A through D) velocity dispersion of $314\,$\kms.

%%%--- Figure 14: Dwarf galaxies ---%%%

We find all galaxies to satisfy our membership constraints. Four of
five spectroscopically detected galaxies lie within the boundary set
by the four main members; galaxy~G, which has the largest
offsets in physical and velocity space, is moving with a radial
velocity offset less than three times the group dispersion. We
therefore consider all five galaxies under consideration here to be members
of HCG~59, based on a strict 3\,$\sigma$-clip, as demonstrated in
Figure~\ref{fig:pspace}.  The inclusion of new members updates the
velocity dispersion of the group to $335~$\kms. Based on this value,
we derive a dynamical mass for HCG~59 of
$M_{dyn}=2.8\times10^{13}~$\Msun, a $\sim10$-fold increase with
respect to the main members alone. This in fact changes the
J07 evolutionary stage of the group from Type~II (intermediate) to Type~III
(gas-poor), as it yields a ratio of H\one-to-dynamical mass of
0.71, with the caveat that the measured H\one\ mass is likely
underestimated because the dwarfs at large group radii are not included.

It is interesting to find dwarf galaxies at large distances from the
center of the group.  This lack of barycentric clustering is also
observed in the Local Group, where it is seen as a morphology-density
relation: dwarf irregulars (dIr) are found at larger distances from
the group center \citep{grebel99} than the quiescent dwarf spheroidals
(dSph) and dwarf ellipticals (dE). This may indicate
that some dIrs are galaxies experiencing their first infalls to the group
center. Such a situation could explain the relatively large velocity
offset and the star-forming nature of galaxy~G.

{The membership of galaxy~G seems the most uncertain of the five galaxies discussed above, given the marginal agreement with the $3\,\sigma$ velocity cut, and the large projected barycentric distance. This is important, as its inclusion does affect the updated dynamical properties, due to the large change in group radius. We quantify this in Table~\ref{tab5}, where we summarize the dynamical properties of HCG~59. Those numbers show a change of mass by a factor of 
$\sim4$ or $\sim10$ with and without galaxy G, while the J07 type changes from II to III regardless of the inclusion of G. The velocity dispersion is most affected: including G increases the value to 336\kms\ from the original 314\kms, while excluding G significantly reduces the dispersion to 208\kms, which is more consistent with the evolved state suggested by our analysis. 
%Its inclusion does not, however, unduly influence the updated dynamical properties, \ie\ the velocity dispersion, dynamical mass, and J07 evolutionary stage.
}

%%%--- TABLE: Dynamics ---%%%
\input{tab5}
%%%

\subsection{The current state of star formation}\label{sec:SF}

We have presented several diagnostics of star formation activity
across HCG~59. The \citet{tzanavaris10} SFRs of galaxies A through D
are $\simeq[4.99, 0.02, 0.16, 0.48]~$\Msun~yr$^{-1}$.  These are
determined from the combination of UV and IR light assuming that all
of the light emanates from star-forming regions.  The presence of
young star clusters in galaxies C~and~D shows that stars are forming
at a fair pace, in accord with the star formation rates quoted in
Table~\ref{tab1}. Overall, the SFRs are consistent with the infrared
SEDs of these galaxies with the notable exception of A which is almost
certainly strongly contaminated by AGN emission and shows no evidence
of ongoing star formation from the other evidence on hand.  There are
several soft X-ray point sources throughout the group, which are
likely to probe compact stellar remnants local to HCG~59
galaxies. There is also soft, diffuse X-ray emission confined to the
galaxies, which probes $\sim10^6~$K gas heated by star formation
(stellar winds and SNe). The IR images do not show much that is
surprising: emission along the spiral arms of~C and in the
star-forming clumps embedded in~D. Four of the five dwarf galaxies
show star formation activity, J
being the exception. They are found to be star-forming based on either
their bright emission lines or blue continua.  Galaxies A~and~B, on
the other hand, exhibit quiescent or even extinguished star formation.
In the case of A, this is inferred by the absence of young star clusters. 

In all, the group does not appear to be undergoing a burst or any
other event notable in terms of current star formation. With the
exception of the two most massive galaxies, the group is forming stars
at a regular pace. This is also exemplified by the behavior of star
cluster complexes across the group, which follow very closely after a
brightness-radius relationship consistent with typical nearby
galaxies, and in contrast to HCGs~7~and~31 \citep{isk10,gallagher10}.

\subsection{Signatures of interactions in the intra-group medium}\label{sec:bplusr}

Major interaction or merger events are very often accompanied by
bursts of star and cluster formation. The examination of star clusters
in HCG~59 presented in Section~\ref{sec:scs} did not show any evidence
of such events in the last few Gyr.  Given the high mass-detection
limit for star clusters at this distance, they cannot be used to trace
minor dynamical events. We therefore search for such evidence in the
lowest surface brightness features detectable in our LCO images. In
Section~\ref{sec:campanas}, we reported the detection of a low surface
brightness stream of material in the projected area between
galaxies~A~and~B. This is visible at low surface brightness in our
\hst\ imaging and quite pronounced in the wide-field images from Las
Campanas, at the $>3\,\sigma$ level. We also detected an arc of
luminous material to the west of galaxy~B, perhaps connecting it to
compact galaxy~I, which we discuss in Section~\ref{sec:dwarfs}.

\begin{figure*}[htbp]
\begin{center}

	\includegraphics[width=\textwidth]{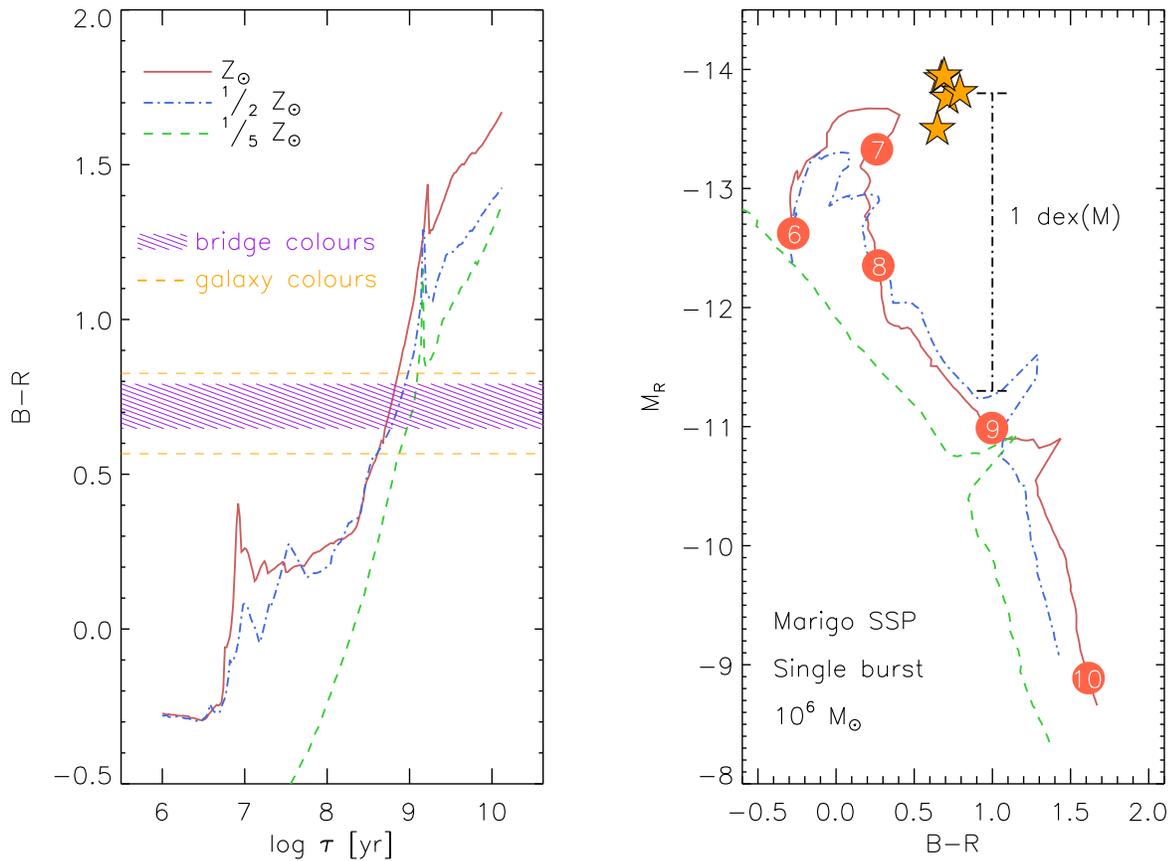}\\
	% bridge.eps
	\caption{Diagnostic plots of the tidal bridge between galaxies A and B. 
		The age-color plot on the {\bf left} shows the evolution of the $B-R$ 
		index over time ($\log\tau$), for \citet{marigo08} models of three 
		different metallicities: $\frac{1}{5}$, $\frac{1}{2}$ and solar, shown 
		as dashed green, dashed-dotted blue and solid red lines respectively. 
		The range of colors in the bridge, as measured in five successive 
		apertures and shown as a shaded region, lines up with the tracks at 
		ages around 1~Gyr (depending on metallicity). 
		The photometry of this faint feature might be affected by diffuse 
		light from the galaxies, which brackets that of the bridge (denoted 
		by dashed yellow lines, galaxy~A is the bottom line). A secondary diagnostic 
		is presented on the {\bf right}, in the form of a color-magnitude 
		diagram. The numbered dots denote log~age in years, while the stars 
		show the color and magnitude of each bridge aperture. If this is a 
		simple stellar population (see text), then we can extrapolate a mass 
		of $\sim10^8~$\Msun\ over the $\sim1~\textup{kpc}^2$ area of the bridge. 
		This measurement can provide constraints for models of the evolution 
		of tidally induced stellar structures. 
	}
	\label{fig:bridge}
	
\end{center}
\end{figure*}

The low surface brightness and limited extent of the `B-I~arc'
preclude precision photometry. We can therefore only pursue an
in-depth analysis of the bridge between~A~and~B. In order to derive
the photometric properties of this feature, we first drew a color-map
to look for an evident \bmr\ gradient. Unfortunately, the stream is
not bright enough to clearly dominate the image background. We thus
conducted photometry of the area and the outskirts of the two galaxy
that bracket it. We used large apertures of radius 75~pc (8~px) to
reduce the background noise.  The results are plotted in
Figure~\ref{fig:bridge}; on the left panel we compare the measured
photometry with the evolution of the \bmr\ color, according to the
\citet{marigo08} model tracks of three metallicities.  We find an intermediate value between the colors of the outskirts of the two galaxies
that define this region (plotted as yellow dashed lines), cautioning
that the emission in this region might be affected by the two galaxy
light-envelopes to some extent.

%%%--- Figure 15: Bridge (A, B), col & CMD ---%%%

The origin of the bridge is not clear and its faintness makes it
difficult to ascertain the dominant source of emission. It could
consist either of stripped stars, or stars that formed {\it in situ} from
stripped gas. If this is mixed stellar material from the two galaxies,
we cannot study it in any more detail. We can, however, develop the {\it in
situ} formation scenario further, by treating the bridge as a simple
stellar population. In this case, the color-magnitude diagram of
Figure~\ref{fig:bridge}~(right), provides an age estimate of about
1~Gyr, depending on metallicity.  In addition, the CMD plotted in this
figure provides an estimate of the stellar mass contained:
with~$10^7~$\Msun\ of~$\sim\,1~$Gyr old stars in each aperture, we
extrapolate a mass in the order of~$\sim10^8$~\Msun, \ie\ a density
of~$\sim100~$\Msun~pc$^{2}$. If the stars here are stripped from a
galaxy, the overall mass will be higher, as the $M/L$ of simple
stellar populations increases with time.

%======= GALAXY B =======
\subsection{HCG~59B as a merger remnant}\label{sec:hcg59b}

HCG~59B, the E/S0 galaxy on the west side of the group, seems
quite regular at first glance, however, a close inspection of the
low-level light reveals some interesting features.
While \textit{ELLIPSE} fitting shows an overall smooth isophotal
structure in \bb\ light, there is severe isophotal twisting in the
central regions. This is spatially coincident with several patches of
extinction we detect in the \hst\ images; they are most pronounced in
\bb, observable in \vb\ and hardly detectable in \ib, implying a thin
column of dust. In the \spit\ bands, the fits indicate very symmetric
structure in the 3.6 and $4.5\,\mu$m bands, but the 5.8 and
$8.0\,\mu$m fits show a faint cusp some $\sim7~$px, or $\sim2.5~$kpc
from the $r^{-1/4}$ surface brightness profile peak. 
% 1.2 asec /px detector scale 
% 7 px = 1.2*7*6e7 / 206265. = 2444 pc

The color-composite IRAC image (Figure~\ref{fig:finder-spit}) shows
hints of structure in galaxy~B; however, an evolved elliptical/lenticular with
near-zero star formation should present a smooth isophotal
profile in all bands. Nonetheless, its SED (Figure~\ref{fig:seds}) shows 
a gradual decline, indicative of emission from stellar photospheres, 
rather than the heated dust associated with star formation. 

To investigate these irregularities further, 
we take advantage of the high resolution of the \hst\ images. 
We construct pixel-by-pixel color maps of B, with the aim of tracing the
exact location of the extinction patches.  If the underlying stellar
population is evolved to the same stage (\ie\ an evolved SSP), then
extinction will be the only source of discrepancies in color. There
are, in fact, three possible sources of color variations in this
filter combination: (i) extinction across a similarly colored stellar
population (mixed or coeval); (ii) spatially separated
stellar populations of various ages and/or metallicities; or (iii) the
presence of gas and star formation -- \ie\ H\two\ region emission
lines.

\begin{figure*}[htbp]
\begin{center}
	
	\includegraphics[width=\textwidth, angle=0]{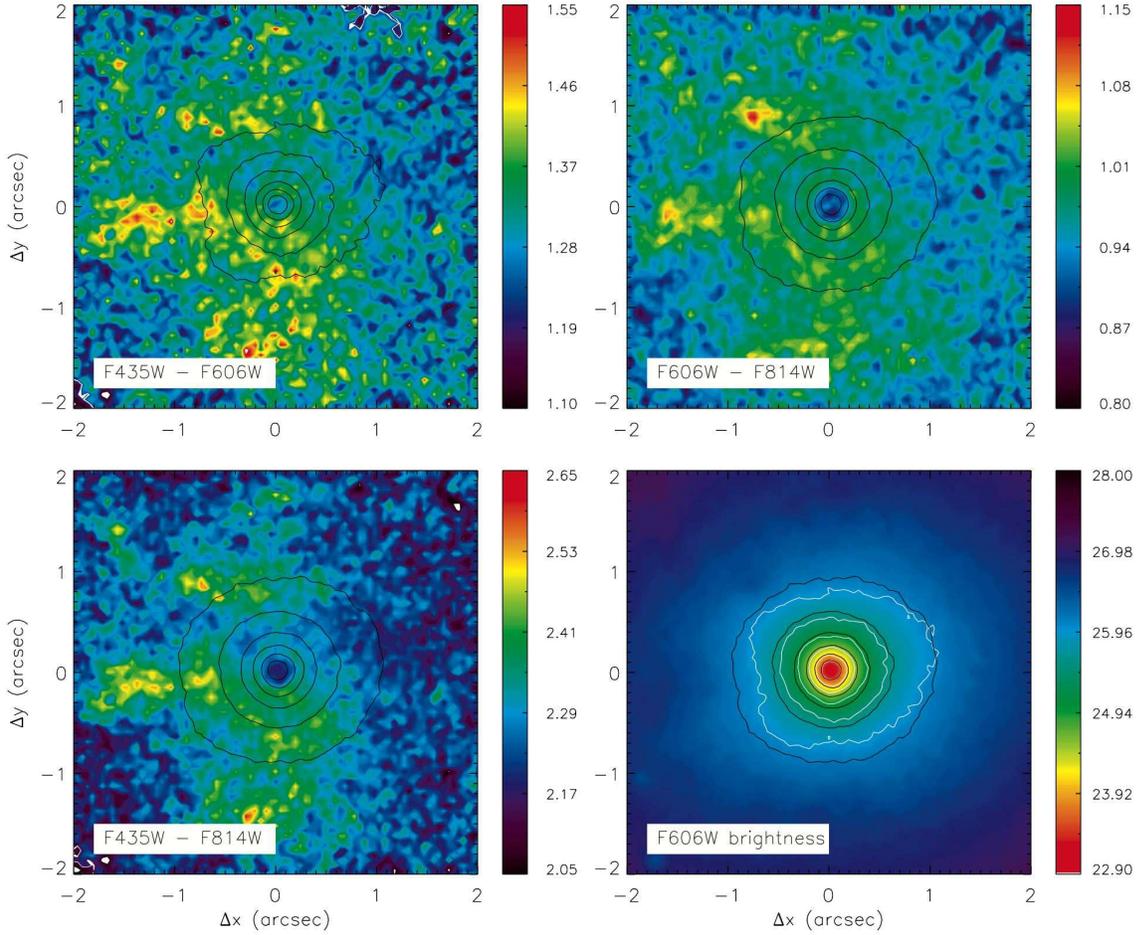}
	% hcg59b.eps
	\caption{\hst\ color maps of the central region of HCG~59B. The maps 
	are centered on the nucleus, defined as the peak of the \ib\ surface 
	brightness profile, and aligned to the world coordinate system, with 
	north to the top and east to the left. The bottom right map shows the 
	\vb\ brightness. Contours show \bb, \vb, \ib, and \ib\ flux in the four 
	panels (upper left, upper right, lower left, lower
	right). Each successive
	contour maps a difference in brightness of 0.5~mag. The $B-I$ map on 
	the bottom left is most sensitive to extinction and does indeed show 
	patches of moderate reddening.  The top row maps also
	show the extinction patches evident in the
	bottom row.  In both the $V-I$ and $B-I$ color maps, the
	nucleus appears bluer than the surroundings; this is
	consistent with additional flux from line emission in the \bb\
	(H$\beta$) and \vb\ (H$\alpha$) filters.  The meaning
	of the observed colors is discussed in more detail in
	Section~\ref{sec:hcg59b}.
	}
	\label{fig:colmapb}
	
\end{center}
\end{figure*}

The maps, shown in Figure~\ref{fig:colmapb}, cover
the three possible filter combinations; for reference, the \vb\ image
is also shown. The $B-I$ map is the one most sensitive to
extinction. To quantify, the \citet{cardelli89} extinction law assigns
almost twice the extinction in the \bb\ as it does to the \ib,
$A_{435}/A_{814} = 1.85$ ($A_{435}/A_V = 1.13$, \cf\ $A_{814}/A_V =
0.61$). This is therefore the map we use to detect patches of low
extinction, of order 0.3~mag, in three fingers extending approximately
eastwards from the north-south line through the nucleus.  This faint
structure is seen in all three colormaps, but not in the \vb\ image.
The \textit{F606W} filter covers various emission lines, including
H$\alpha$, [N\two], [S\two], H$\beta$, and [O\three]. Interestingly,
\citet{martinez10} found H$\alpha$, [NII] and [SII] emission in the
spectrum of B, at relative intensities consistent with a composite
H\two\ region plus AGN emission. From the concentrated, blue core of
the $V-I$ image, this line emission appears to be spatially
coincident with the nucleus.

%%%--- Figure 16: Color map B ---%%%

Circumstantial evidence for a close encounter in the recent history of
galaxy~B is provided by the uneven distribution of its GC system. GCs
normally form spherical haloes, however, here we find GCs at large
radii, many concentrated along the stellar stream that seemingly
connects galaxies A~and~B, and its extension across the far side of
galaxy~B.  Furthermore, the overabundance of GCs in galaxy~B is
matched by a severe dearth of clusters in A. Given the possibility
that the two galaxies interacted $\sim1$ Gyr ago, a scenario whereby
GCs are transferred between the two systems is not out of the
question. It is unclear from a dynamical perspective why in the
process of an interaction the GCs would flow from A, the more massive
entity, to B. In a simple thought experiment, we move as many clusters
from B to A as are required to level the $S_N$ of A to the nominal
value for an Sa. This still leaves an excess of GCCs in B relative to
normal.  However, the factor-of-two uncertainties involved in the
determination of $S_N$ do not rule out this scenario of GC `swapping'.

The X-ray map of this galaxy, as described in Section~\ref{sec:xrays},
reveals two distinct X-ray sources in the nuclear region. 
Unfortunately, with full-band luminosities of  $L_X = (1.4, 1.7)\times10^{39}~$erg~s$^{-1}$, neither is sufficiently luminous be identified as an unambiguous AGN -- a possibility that the optical spectroscopy of
\citeauthor{martinez10} has indicated.  Given
the lack of ongoing star formation in the region, the sources are
unlikely to be high mass X-ray binaries, though luminous low mass
X-ray binaries (associated with older stellar populations) or groups
of them unresolved at the distance of HCG~59 are plausible.  In addition, due to the
uncertainty in matching X-ray sources to optical imaging, we cannot
confidently derive a one-to-one correlation between the optical clumps
and these sources, although the correlation is confirmed to within the
X-ray positional uncertainties.  We note however that, as discussed in
Section 2.7, the nuclear X-ray sources are spatially close to the
optical center of the galaxy.

The excess of globular clusters and uneven dust distribution in the
nuclear regions of B hint at some interaction in the more distant
($>$~Gyr) past, but the lack of additional evidence for structural
disturbances limits our ability to infer more.  We do detect a few
young clusters in galaxy~B and a non-zero (though low) SFR, and
therefore some reservoir of cold gas is present.  Accretion of a
satellite galaxy is therefore a possibility.  Furthermore, the unimodal GC
color distribution does not favor a gas-rich, major merger in the
past.  This conclusion follows the paradigm of \citet{muratov10}, who
attribute the known bimodality of GC colors to late-epoch mergers.  It
is also a reasonable assumption that the implied interaction did not
feature a major merger with a gas-rich system, as that would have enhanced
the young and intermediate-age cluster populations.

%======= HCG 59A =======
\subsection{Nuclear activity in HCG~59A}\label{sec:hcg59a}

In Section~\ref{sec:obs-spit} we reported that the IR emission in
galaxy~A is consistent with being dominated by an AGN, rather than
star formation. This is based on the disparity between the galaxy's
morphological type of Sa and the high UV+IR SFR suggested by
interpreting the emission as related to SF. Furthermore, the lack of
young massive clusters is inconsistent with a SFR of
5~\Msun$\,\textup{yr}^{-1}$.

This hypothesis is supported further by the finding of a hard X-ray
source in the nuclear region of galaxy~A, with $L_X =
1.1\times10^{40}~$erg~s$^{-1}$ as reported in
Section~\ref{sec:xrays}. The spectroscopic AGN survey of HCGs
\citep{martinez10} places the galaxy at the interface of the H\two\
and AGN zones in the `BPT' diagram \citep{bpt}, based on optical
emission-line ratios. The 2\asec-wide slit they used encompasses a
wide region (effective aperture of 0.58~kpc), therefore the signal is
most likely diluted by circumnuclear and disk light.  Visual
inspection of the central region reveals asymmetric structure,
resembling a second nuclear source of comparable \ib\ luminosity to
the nucleus. As in the previous section, we employ \hst\ color maps to
take advantage of the spatial resolution of $\simeq12~$pc per pixel.
% 2" @ 60 Mpc = 582 pc

\begin{figure*}[htbp]
\begin{center}

	\includegraphics[width=\textwidth, angle=0]{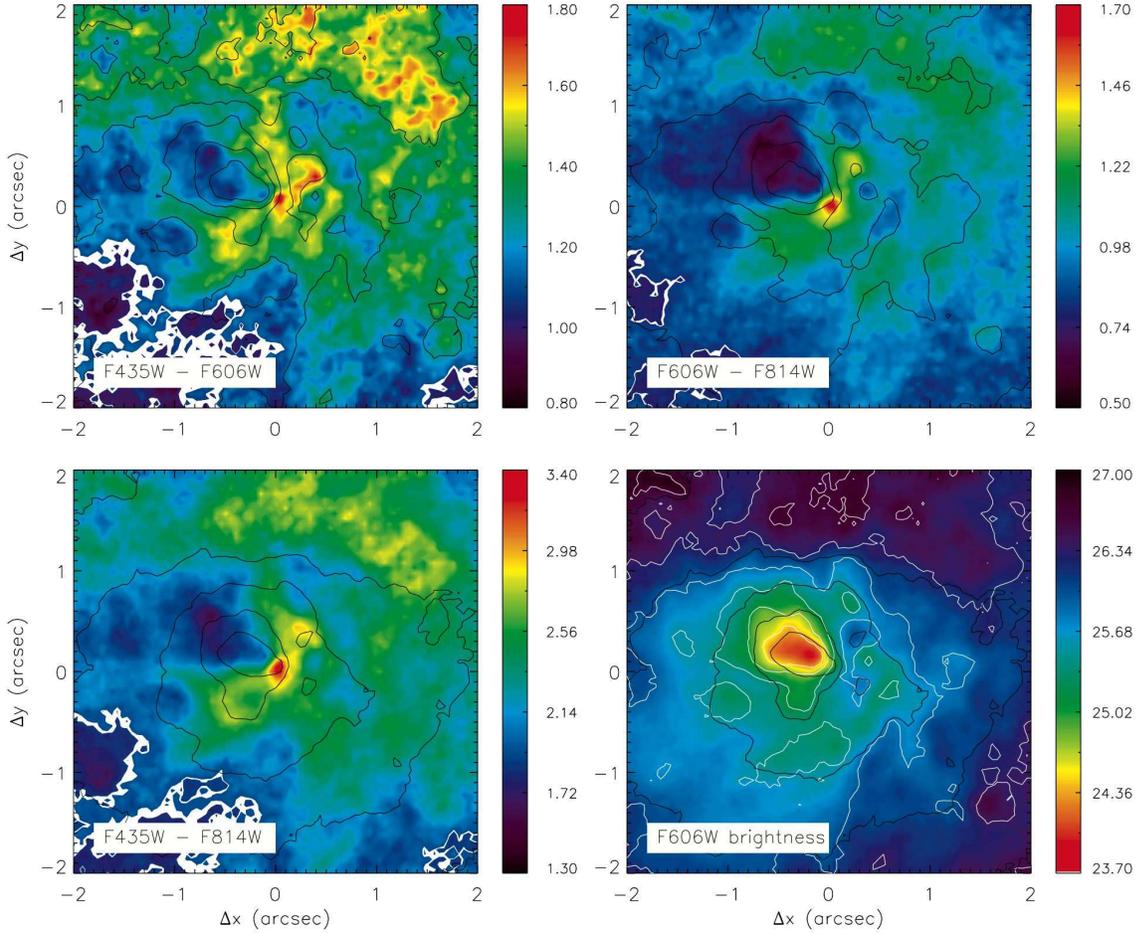}
	% hcg59a.eps
	\caption{\hst\ color maps of the nuclear region of HCG~59A, following
		the plotting technique presented in Figure~\ref{fig:colmapb}. North 
		is to the top and east is to the left. The central peak appears 
		elongated in the \vb\ flux map (lower right) and shows a composite 
		structure with two additional peaks nearby and a long tail. The tail 
		could be an artifact introduced by the complex and heavy extinction in 
		this region, perhaps caused by the inner spiral structure of HCG~59A. 
		It could alternatively be associated with the nuclear activity we 
		note in the X-ray image of this galaxy (Section~\ref{sec:xrays}), as 
		well as the significant excess in 8 and $24~\mu$m emission, discussed in 
		Section~\ref{sec:obs-spit}. A blue cone-shaped structure is most 
		evident in the \vb--\ib\ color map (top right). The nature of this 
		blue cone is discussed in Section~\ref{sec:hcg59a}. 
	}
	\label{fig:colmap}
	
\end{center}
\end{figure*}

The maps, shown in Figure~\ref{fig:colmap}, reveal a cone of blue
light at $(x, y) \simeq (-0.1, 0.1)$, with colors of
$B_{435}-V_{606}\simeq0.9$, $V_{606}-I_{814}\simeq0.6$, and
$B_{435}-I_{814}\simeq1.6$.  The complexity of the central region
inhibits easy interpretation, but one clue are the emission lines covered in the
three bands. \vb\ covers H$\beta$, [O\three], H$\alpha$ and [N\two],
which can be associated with star formation and/or AGN
activity.  In this scenario, the blue cone could stand out as a result
of geometry, perhaps being located in a break in the dust
distribution. A simple explanation could relate this feature to an
unreddened line of sight through the inner spiral structure of A. In
that case, however, we would expect to see bright, young star
clusters, as they are known to shine through thick columns of dust,
let alone relatively dust-free regions \citep[\eg\ Region~B in
M82;][]{smith07regb,isk08}.

%%%--- Figure 17: Color map A ---%%%

A different interpretation can relate this structure to an AGN.  The
lack of symmetry could suggest a small narrow-line region,
photoionized by the AGN continuum, with projection effects and
obscuration hiding the cone on the far side.  This would produce
strong [O\three] and [N\two] emission, the presence of which was
reported by \citet{martinez10}. This geometry is consistent with the
inclination of galaxy~A of no more than $30^{\circ}$ (assuming the AGN
and galaxy share the same inclination angle).

Combining the pieces of evidence collected from the X-ray, optical and
MIR emission, we propose that the nuclear emission in HCG~59A is
dominated by a low-luminosity AGN with a photoionized narrow-line region.  The
onset of activity may be related to a possible encounter with galaxy~B
about~1~Gyr~ago, as tentatively dated from the colors of the bridge
connecting the two galaxies (Section~\ref{sec:bplusr}).

%======= SUMMARY =======
\section{Summary}\label{sec:summary}

We have presented an analysis of HCG~59, a compact group comprising
four main galaxies and at least five newly discovered dwarfs at the $M_r<-15.0~$mag
level. Our results are based on multi-wavelength observations and
continue a series of papers that have followed two different
approaches: on the one hand we have treated the overall properties of
Hickson compact groups
\citep{johnson07,gallagher08,tzanavaris10,walker10}; on the other
hand, we have surrounded our \hst\ observations with a
multi-wavelength dataset to pursue in-depth investigations of
individual CGs, one at a time \citep{palma02,gallagher10,isk10}.

Compared to HCGs~7 and 31, two compact groups previously studied in
this series, HCG~59 presents something of an intermediate step: where
HCG~7 was found to be interacting solely in the dynamical sense (\ie\
currently in the absence of direct hit encounters), HCG~59 shows
evidence for stronger interactions in the recent past.  There is
evidence for star formation in the intragroup medium in the H~{\sc ii}
regions to the Northwest of galaxy B, in sharp contrast to HCG~7 where
the star-formation associated with each galaxy was self-contained.  In
the context of the evolutionary sequence we proposed in \citet{isk10},
it occupies a stage further along than HCG~7.  It has begun building
an IGM, as testified by some amount of intra-group light, but has
yet to build up a large elliptical fraction. Its classification as a
relatively unevolved group (J07) is in accord with standard diagnostics 
such as its velocity dispersion of $\sim335~$\kms\ and lack of diffuse,
extended X-ray emission.

Through the use of SDSS data, we associated five dwarf galaxies
to~HCG~59 for the first time. Their inclusion updated the velocity
dispersion and dynamical mass of the group and changed its
J07 evolutionary type to an evolved group of Type~III (originally 
Type~II, or intermediate). 
The star-forming nature of these dwarfs, their radial velocities, and
distances from the group core seem to suggest that some may be
infalling for the first time. The noted relation between morphological
type and barycentric distance follows the one observed in the Local
Group and may be considered -- conversely -- as an indication of the
`young' dynamical state of HCG~59. The above information highlights the
importance of studying the dwarf galaxy contingent of compact groups.

Star formation is proceeding at a regular pace in this CG, certainly
at a rate consistent with the morphologies of the member galaxies. The
regularity of star cluster complexes agrees with this image. The star
cluster population does not show evidence of major, gas-rich,
interactions in the past $\sim \textup{Gyr}$ and the IR SEDs are
generally as expected. One exception to this rule is galaxy~D. 
Given its large size, the morphological
regularity of its neighbors, and the lack of evidence of recent interactions, it is not clear why it has such an irregular structure and why it is forming
stars at the rate that it is.
 
Where the information is unclear for galaxy~D, B is evidently in the
midst of at least one dynamical process. 
This probably started more than a Gyr
ago given the lack of tidal features such as shells and tails
commonly observed in such events in the optical
\citep{schweizer98,rogues,mullan11}.  
It is likely physically associated with an arc of star formation that
seemingly connects it to dwarf galaxy~I (the arc that hosts the discovered extragalactic H\two\ regions), and 59B lies at one end of a
low surface brightness bridge of stellar material, which might
physically connect it with A. We found this feature to be no older
than $\sim1~$Gyr by age-dating its stellar population.  Perhaps most
intriguingly, the globular cluster population of B is anomalously
large, with significantly more globular cluster candidates than A,
despite its lower stellar mass.  The origin of this discrepancy is
unclear, but hints at an additional event in the more distant past.

In one sense, this group is typical of early or intermediate-stage
HCGs. There are plenty of dynamical processes at play, however they
are all proceeding at a low level and centered around one or two
objects. Galaxy~B is at the focus of all such processes that our
diagnostics can reach, like galaxy~B in HCG~7 (also an early-type
galaxy). In addition, both these groups may feature an active nucleus
host galaxy (the dominant galaxy~A in both groups, albeit the
detection is not certain in HCG~7).

We find the presence of a low-luminosity AGN in A to be likely given
the spatial coincidence of the galaxy centroid with a
$\sim10^{40}$~\lumin\ X-ray point source.  The inferred SFR from the
IR+UV luminosity is not supported given the lack of young star clusters,
and therefore likely is overestimated because of significant AGN
contamination.  If the A-B bridge constitutes a physical connection
between the two galaxies, a causal connection between the interaction
and the AGN is possible. In this case, we can constrain the timescale
since the onset to no more than one Gyr. 

In the introduction, we discussed HCGs as a potentially special
environment in terms of galaxy evolution. The apparent duality of
`modes' in which HCG galaxies are found -- either star-forming or quiescent --
and the evident lack of an intermediate stage population are in accord with the
mid-IR color-space `gap' discussed in our previous work
\citep{johnson07,tzanavaris10,walker10}. The impact of the group
environment will be the topic of the next paper in this series. There
we will treat HCGs~16,~22~and~42 with a goal of understanding the
evolutionary processes at play in compact groups and relating their
galaxies to those found at other levels of galaxy clustering.

\section*{Acknowledgements}
	We thank the referee, Cristiano Da Rocha, for his constructive 
	criticism of the manuscript and suggested additions that
	elevated the work. 
	ISK thanks Ranjan Vasudevan and Matt Povich for educational
	discussions on the X-ray properties of AGN
	and star-forming regions. We thank Gordon Garmire for his
	contribution in obtaining the X-ray dataset.
	Support for this work was provided by NASA through grant
	number HST-GO-10787.15-A from
	the Space Telescope Science Institute which is operated by
	AURA, Inc., under NASA contract
	NAS 5-26555 and through Chandra Award No. GO8--91248 issued by
	the Chandra X-ray Observatory Center, which is operated by the
	Smithsonian Astrophysical Observatory under NASA contract
	NAS8-03060.  SCG, KF, and ARH thank the National Science and
	Engineering Council
	of Canada and the Ontario Early Researcher program. Funding
	was provided by the National Science Foundation under award 0908984. 
	PRD would like to acknowledge support from \hst\ grant HST-GO-10787.07-A.
	AIZ acknowledges support from the NASA Astrophysics Data Analysis Program
	through grant NNX10AE88G. 
	KEJ gratefully acknowledges support for this work provided by
	NSF through CAREER award
	0548103 and the David and Lucile Packard Foundation through a Packard Fellowship.
        PT acknowledges support through a NASA Postdoctoral Program Fellowship
        at Goddard Space Flight Center, administered by Oak 
        Ridge Associated Universities through a contract with NASA.
	This research has made use of the NASA/IPAC Extragalactic
	Database (NED) which is operated by
	the Jet Propulsion Laboratory, California Institute of
	Technology, under contract with the National
	Aeronautics and Space Administration. 

{\it Facilities:} \facility{HST ()}, \facility{Spitzer ()}, \facility{Chandra ()}, \facility{Las Campanas ()}

\bibliographystyle{apj}
\bibliography{references}

\end{document}

%% file: tab1.tex
%%%--- TABLE 1: BASIC INFO ---%%%
\begin{table*}[htbp]
\begin{center}

\caption{Basic information on HCG~59 main members}\label{tab1}
\begin{tabular}{lccccr} 
\tableline
\tableline
Identifier		& % C1: ID
Coordinates\tablenotemark{a}	& % C2: Coordinates
{Type}	& % C3 C4: Morph
$m$			& % C5: mag
$v_R$			& % C6: RV
References\tablenotemark{b}	\\ % C7: Refs
			& % C1
(J2000)			& % C2
H89\tablenotemark{c} 	& % C3
(mag) 			& % C5
(\kms) 			& % C6
			\\ % C7
\tableline
A: IC~0737		& 11:48:27.55 +12:43:38.7	& Sa					& 14.82 ($B$)	& 4109  & [1], [2], [3]\\ 
B: IC~0736		& 11:48:20.08 +12:42:59.5	& E0\tablenotemark{d}	& 15.60 ($B$)	& 4004  & [4], [2], [5]\\
C: KUG~1145+129	& 11:48:32.44 +12:42:19.5	& Sc					& 15.90 ($g$) 	& 4394  & [4], [2] \\
D: KUG~1145+130	& 11:48:30.64 +12:43:47.8	& Im					& 16.00 ($g$)	& 3635  & [4], [2]\\
%
%coordinates from Allison - centroids from GALFIT of HST images:
%59a: 11:48:27.549, +12:43:38.72
%59b: 11:48:20.075, +12:42:59.47
%59c: 11:48:32.441, +12:42:19.48
%59d: 11:48:30.635, +12:43:47.76

\tableline
\end{tabular}
\tablenotetext{a}{Coordinates are the centroids from fitting the \hst\ \ib-band images of each galaxy with S\'ersic profiles.  See \S~\ref{sec:obs-hst} for more details.} 
\tablenotetext{b}{[1]:~\citet{chandracat}; [2]:~\citet{rc3}; [3]:~\citet{hickson92}; [4]:~\citet{sdss}; [5]:~\citet{zwickycat}.}
\tablenotetext{c}{\citet{hickson89}}.
\tablenotetext{d}{The RC3 designation for 59B is` S0?', and so there is some uncertainty as to its classification.}
\end{center}
\end{table*}

%% file: tab2.tex
%%%--- TABLE 2: MASS, SFR, NUCLEI ---%%%
\begin{table*}[htbp]
\begin{center}

\caption{Masses and star formation rates of HCG~59 galaxies}\label{tab2}

\begin{tabular}{l r@{.}l r@{.}l c r@{.}l l} 
	
	\tableline
	\tableline
	ID															& % C1: -
	\tmult{$M*$\tablenotemark{a}}								& % C2: Mstar
	\tmult{$M_{\scriptsize \textup{H}_2}$\tablenotemark{b}}	& % C3: M_H2
	SFR\tablenotemark{c}										& % C4: SFR
	\tmult{sSFR\tablenotemark{d}}								& % C5: SSFR
	Nucleus \tablenotemark{e}									\\ % C6: Nuclear activity
													& % C1
	\multicolumn{4}{c}{($\times10^{9}~$\Msun)}		& % C2 & C3
	(\Msun~yr$^{-1}$)								& % C4
	\tmult{($\times10^{-11}~\textup{yr}^{-1}$)}		& % C5
													\\ % C6
	\tableline
	A & 17&40	& 10&2		& $4.99\pm0.67$\tablenotemark{f}	& \hspace{20pt}28&66 	& 	C/AGN\\
	B & 8&29	& $<8$&8	& $0.02\pm0.01$						&  			0&19		& 	C\\
	C & 3&03 	& $<7$&2	& $0.16\pm0.03$						&  			5&15		& 	H\two\\
	D & 2&67 	& $<6$&6	& $0.48\pm0.04$						& 			18&15		& 	H\two\\
\tableline
\end{tabular}
%%%%%
\tablenotetext{a,c,d}{Stellar masses, star formation rates (SFR) and specific SFRs (sSFR) are drawn from 
	\citet{tzanavaris10}.  The published stellar masses were off by a factor of 7.4; these values have been corrected for this error.}
\tablenotetext{b}{From the CO observations of \citet{vm98}}
\tablenotetext{e}{From \citet{martinez10}; `C' stands for `composite', \ie\ one that falls at the
	H\two /AGN overlap region, as defined in \citet{kewley06}; see Sections~\ref{sec:hcg59b}, 
	\ref{sec:hcg59a} for discussion on these designations. }
\tablenotetext{f}{This value is heavily affected by the AGN in galaxy A, as will be elaborated in Section~\ref{sec:obs-spit}.}

\end{center}
\end{table*}

%% file: tab3.tex
%%%--- TABLE: DWARF GALAXIES ---%%%
%SCG: updated and reordered 2011 Jul 13
%\begin{sidewaystable}[htbp]
\begin{table*}[htbp]
\begin{center}

\caption{Dwarf galaxies in HCG~59}\label{tab:dwarfs}

\begin{tabular}{lcccccccc}
	
\tableline
\tableline
ID					& % C1: HCG ID
SDSS ID				& % C2: SDSS ID
(Plate, MJD, Fiber)			& % C3: 
Morphology			& % C4: Morph type
$r$\tablenotemark{a}				& % C5: r
$A_r$\tablenotemark{b}						& % C6: extinction
$M_r$\tablenotemark{c}				& % C6: AbsRmag
$v_R$\tablenotemark{b}				& % C7: Radial velocity
$d_\textup{\scriptsize GC}$	\\ % C8: projected barycentric distance
									& % C1
									& % C2
									& % C3
									& % C4
\multicolumn{3}{c}{(mag)}			& % C5, 6, 7
(\kms)								& % C8
(kpc)								\\ % C9
\tableline

F & J114930.72+124037.5 & (1609, 53142, 464) &  Sd & 15.83 &  0.09 & $-18.15$ & 3984 &   290\\
G & J114940.11+122338.6 & (1609, 53142, 465) & dIr & 17.34 &  0.08 & $-16.63$ & 3195 &   471\\
H & J114912.21+123753.8 & (1608, 53138, 625) & dIr & 17.39 &  0.08 & $-16.58$ & 4011 &   225\\
I & J114817.89+124333.1 & (1608, 53138, 586) & dIm & 17.43 &  0.10 & $-16.56$ & 4038 &    32\\
J & J114813.50+123919.2 & (1608, 53138, 593) &  dE & 17.69 &  0.10 & $-16.30$ & 3942 &    83\\
%
% m-M = 33.89
%
% 
%center position for SDSS search
%177.11086    12.71123
%E(B-V)=0.037
\tableline
\end{tabular}
\tablenotetext{a}{SDSS model $r$ magnitudes.  These values have not been corrected for Galactic extinction.}  
\tablenotetext{b}{Galactic extinction and recession velocities are from SDSS DR7.}
\tablenotetext{c}{Absolute $r$ magnitudes assuming a distance modulus of 33.89 and correcting for Galactic extinction.}
\end{center}
%\end{sidewaystable}
\end{table*}

%% file: tab4.tex
\begin{table*}[htbp]
\begin{center}

\caption{Observed and derived properties of the globular cluster systems in HCG~59}\label{tab-gc}

\begin{tabular}{lr@{$\pm$}lr@{$\pm$}lr@{$\pm$}lr@{$\pm$}lr@{$\pm$}l} %{lr@{\pm}lrrr}
	
\tableline
\tableline

Galaxy													& % C1: Galaxy ID
\tmult{$N_\textup{\scriptsize GCC}$\tablenotemark{a}}	& % C2: N_GC(detected)
\tmult{$N_\textup{\scriptsize back}$\tablenotemark{b}}	& % C3: BG sources
\tmult{$N_{V<26}$\tablenotemark{c}}						& % C4: N_GC with V < 26 mag
\tmult{$N_\textup{\scriptsize total}$\tablenotemark{d}}& % C5: N_total, derived
\tmult{$S_N$ \tablenotemark{e}}							\\% C6: Specific frequency
\tableline
%
%c1   c2 err       c3  err          c4  err        c5  err        c6   err
 A &  18&4	 & 6&1.0	&  13&6		&  33&21	& 0.3&0.2\\
 B & 162&13	 & 6&1.0	& 180&34	& 462&182	& 7.7&3.0\\
 C &   7&3	 & 6&1.0	&	1&3		&   3&8	& 0.1&0.3\\   
 D &  11&3	 & 1&0.2	&  11&5		&  28&18	& 1.6&1.0\\
\tableline
\end{tabular}
\tablenotetext{a}{Number of detected GCCs}
\tablenotetext{b}{Background correction}
\tablenotetext{c}{Number of clusters brighter than $V=26~$mag}
\tablenotetext{d}{Estimated total GC population}
\tablenotetext{e}{Specific frequency}
\end{center}
\end{table*}

%% file: tab5.tex
\begin{table*}[htbp]
\begin{center}

\caption{Updated dynamical properties of HCG~59}\label{tab5}

\begin{tabular}{lcccc}
	
\tableline
\tableline

Reference			& % C1: Reference
$\bar{v}$			& % C2: Mean velocity
$\sigma_v$			& % C3: Velocity dispersion
$M_{dyn}$			& % C4: Dynamical mass
J07 type			\\ % C5: Johnson 07 type
					& % C1
\tmult{(\kms)}		& % C2, 3
(\Msun)				& % C4
					\\ % C5
\tableline
\citet{hickson82}		& 4036 & 314 & $2.9\times10^{12}$ & II\\
This work				& 3924 & 336 & $2.8\times10^{13}$ & III\\
This work, excluding G	& 4015 & 208 & $1.1\times10^{13}$ & III\\
\tableline
\end{tabular}
%   
%\tablenotetext{a}{Number of detected GCCs}
%
% Giants only:	4035.50      313.750
% With G:		3923.56      335.592
% Without G:	4014.62      208.338
% 
% 
\end{center}
\end{table*}